\documentclass[aps,prb,twocolumn,amsmath,amssymb,nofootinbib,eqsecnum,superscriptaddress,floatfix,preprintnumbers]{revtex4}

\usepackage{amsmath}
\usepackage{amssymb}
\usepackage{amsthm}
\usepackage[dvips]{color} 
\usepackage{graphicx}
\usepackage{dcolumn} 
\usepackage{bm} 
\usepackage[hypertex]{hyperref}
\usepackage{longtable}
\usepackage{ulem}   
\renewcommand\d{\partial} 
\newcommand\x{{\bm{x}}} 
\newcommand\p{{\bm{p}}} 

\begin{document}

\preprint{MIT-CTP 4186}

\title{
Counting Majorana zero modes in superconductors
      }

\author{Luiz Santos} 
\affiliation{
Department of Physics, 
Harvard University, 
17 Oxford Street, 
Cambridge, Massachusetts 02138,
USA
            } 

\author{Yusuke Nishida} 
\affiliation{Center for Theoretical Physics, Massachusetts Institute of Technology, Cambridge, Massachusetts 02139, USA}

\author{Claudio Chamon} 
\affiliation{
Physics Department, 
Boston University, 
Boston, Massachusetts 02215, USA
            } 
\author{Christopher Mudry} 
\affiliation{
Condensed matter theory group, 
Paul Scherrer Institute, CH-5232 Villigen PSI,
Switzerland
            } 

\date{\today}

\begin{abstract}
A counting formula for computing the number of 
(Majorana) zero modes bound 
to topological point defects is evaluated in
a gradient expansion for systems with 
charge-conjugation symmetry. This semi-classical
counting of zero modes is applied to 
some examples that include graphene and 
a chiral $\mathsf{p}$-wave superconductor
in two-dimensional space. In all cases,
we explicitly relate the counting of zero modes
to Chern numbers.
\end{abstract}
\maketitle

\section{
Introduction
        }
\label{sec: Introduction}

The counting of zero modes, 
eigenstates annihilated by a single-particle Hamiltonian 
$\mathcal{H}$,
has a long history in physics. Charge-conjugation symmetry,
the existence of a norm-preserving linear (antilinear)
transformation $\mathcal{C}$
that anticommutes with $\mathcal{H}$, protects the parity 
of the number of zero modes. 
When the parity is odd, at least one zero mode must
be robust to any perturbation that preserves the charge-conjugation
symmetry. This paper aims at calculating the parity of zero modes
of a single-particle Hamiltonian
$\mathcal{H} \big( \hat{\boldsymbol{p}},
\boldsymbol{\varphi}(\boldsymbol{x}) \big)$
when (1) it obeys charge-conjugation symmetry, 
(2) it describes fermionic quasiparticles, 
and (3)
it depends on a position dependent vector-valued order parameter
$\boldsymbol{\varphi}(\boldsymbol{x})$.
These three assumptions are often met in mean-field treatments
of electrons interacting with each other or with collective
excitations such as phonons or magnons
in condensed matter physics.

Zero mode solutions can be found by direct means, 
in practice solving a differential equation.
This requires a non-universal definition of the model since
both microscopic and macroscopic data must be supplied, 
say the boundary conditions to be obeyed at the origin 
and at infinity in space.

Is there an alternative approach to calculating the parity in the number of
zero modes that is more universal? The celebrated index theorem for
elliptic differential operators gives a positive answer to this question
for those problems in physics for which this theorem applies (Dirac Hamiltonians
for example).%
~\cite{index thm} 
The index theorem achieves this by relating some (not all!) 
zero modes to a topological number (a global property of the Hamiltonian 
that can only take integer values). However, the index theorem
cannot be applied to most Hamiltonians of relevance to condensed matter
physics.

In this paper, we start from an exact integral representation
of the total number of \textit{unoccupied} zero modes 
(up to exponential accuracy) $N$ 
for fermionic single-particle Hamiltonians with charge-conjugation symmetry 
in terms of a conserved quasiparticle charge $Q$,
\begin{equation}
N=-2Q.
\label{eq:  Schrieffer counting formula from intro}
\end{equation}

\begin{figure}
\includegraphics[angle=0,scale=0.5]{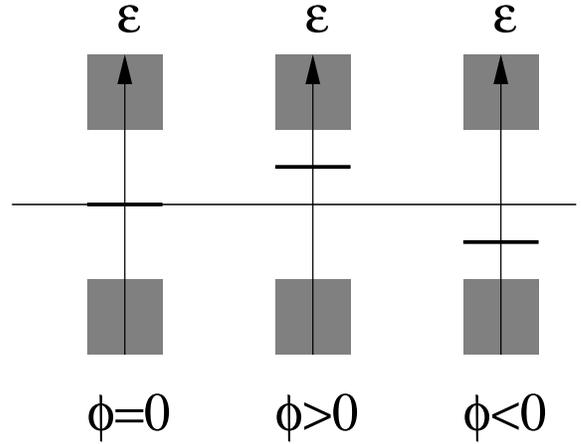}
\caption{
The charge-conjugation-symmetry-breaking parameter $\phi$
moves in energy a mid-gap state upward or downward depending 
on its sign. The continuum parts of the energy-eigenvalue
spectrum are denoted by the shaded boxes. The thin horizontal
line denotes the band center about which the spectrum is
symmetric when $\phi=0$.
        }
\label{Fig: energy levels}
\end{figure}

This counting formula appears implicitly in Ref.~\onlinecite{Su81} and 
explicitly in Ref.~\onlinecite{Jackiw81a}, 
both in the context of polyacetylene.%
~\cite{Wilczek02} 
For polyacetylene, it relates the number of unoccupied zero modes
to the \textit{conserved electric charge} 
$Q^{\,}_{\mathrm{dw}}$ 
induced by domain walls
in the spontaneous bond ordering triggered by the coupling of electrons
to phonons. Remarkably, the electric charge 
$Q^{\,}_{\mathrm{dw}}=\pm1/2$ induced by a single domain wall
is fractional and counts a single zero mode, with the sign ambiguity
resolved by whether the midgap state is filled or empty.%
~\cite{Jackiw76,Su79,Su80}
Alternatively, this sign ambiguity can be removed 
by the application of a small charge-conjugation-symmetry-breaking
perturbation that shifts the energy of the zero mode up or down
(see Fig.~\ref{Fig: energy levels}). Hence, the counting formula%
~(\ref{eq:  Schrieffer counting formula from intro})
becomes
\begin{equation}
N=2Q\hbox{ mod }2
\label{eq:  Schrieffer counting formula from intro bis}
\end{equation}
if no prescription is given as to whether the filled Fermi sea includes 
or not a zero mode. The same assignment of quantum numbers also relates
a single zero mode and the conserved electric charge 
$Q^{\,}_{\mathrm{Kekule}}$
induced by a vortex with unit vorticity
in the Kekul\'e dimerization pattern
of graphene.%
~\cite{Hou07,Jackiw07,Chamon08a,Chamon08b,Ryu09} 

For polyacetylene and graphene, the charge-conjugation symmetry
is approximate, for it originates from a sublattice symmetry
that is broken as soon as next-nearest-neighbor hopping
is included in the tight-binding model. To the extend
that the Bardeen-Cooper-Schrieffer (BCS)
mean-field approximation to superconductivity
is empirically observed to be excellent,
single-particle Bogoliubov-de-Gennes (BdG) Hamiltonians
realize a much more robust charge-conjugation symmetry.%
~\cite{footnote: Andreev bound states}
For BdG Hamiltonians, 
zero modes are associated to Majorana fermions.
Majorana fermions do not carry a well-defined electric charge
since global electro-magnetic gauge invariance is broken
in any mean-field treatment of superconductivity.
The counting formula%
~(\ref{eq:  Schrieffer counting formula from intro})
nevertheless applies to any
BdG Hamiltonian $\mathcal{H}^{\,}_{\mathrm{BdG}}$
with the important caveat that 
the conserved quasiparticle charge
$Q^{\,}_{\mathrm{BdG}}$ 
is unrelated to the electric charge.
Rather, the conservation of $Q^{\,}_{\mathrm{BdG}}$ 
encodes, for any local BdG Hamiltonian, 
a local continuity equation
obeyed by the Bogoliubov quasiparticles
that is responsible for the conservation of
the thermal flow in the mean-field treatment of 
superconductivity.%
~\cite{Altland97,Senthil98,Senthil00}

In this paper, we represent the counting formula%
~(\ref{eq:  Schrieffer counting formula from intro})
in terms of single-particle Green functions. 
The advantage of this choice is that it easily 
lends itself to a perturbative (gradient) expansion
of the conserved quasiparticle charge $Q$
that forgoes the non-universal short-distance data.%
~\cite{Niemi86,Volovik91,Volovik:book}
To leading order, this expansion is akin to 
an adiabatic approximation.
We thus propose the adiabatic approximation to
the conserved quasiparticle charge $Q$ 
as an efficient mean to compute the parity in the
number $N$ of unoccupied zero modes of
charge-conjugation-symmetric single-particle Hamiltonians.

We then apply this formula to charge-conjugation-symmetric Dirac
insulators with time-reversal symmetry in arbitrary dimensions and
chiral $\mathsf{p}$-wave BdG superconductor in
two-dimensional space that all support a point defect.
One of the main results of this paper is an integral representation
for the number (zero or one)
of unoccupied zero mode induced by a unit
vortex in a two-dimensional chiral 
$\mathsf{p}$-wave BdG superconductor.
A by-product of this integral representation 
is that it is closely related to \textit{the second Chern number}.
We also relate \textit{the $d$-th Chern number}
to the number of unoccupied zero modes
for Dirac fermions in $d$-dimensional
space that interact with a $d$-tuplet of Higgs field
supporting point defects.

The gradient expansion is presented in 
Sec.~\ref{sec: Gradient expansion}
and applied to point defects
in Sec.~\ref{sec: zero modes induced by point defects}.
We conclude in Sec.~\ref{sec: Conclusion}.

\section{
Gradient expansion of the counting formula
        }
\label{sec: Gradient expansion}

Let the charge-conjugation-symmetric single-particle Hamiltonian 
$\mathcal{H}
\big(
\hat{\boldsymbol{p}},
 \boldsymbol{\varphi}(\boldsymbol{x})
\big)$
be such that its dependence on the space coordinate 
\begin{equation}
\boldsymbol{x}\in\mathbb{R}^{d}
\label{eq: def space}
\end{equation}
is implicit through that of a static vector-valued order parameter
\begin{equation}
\boldsymbol{\varphi}(\boldsymbol{x})\in\mathbb{R}^{D},
\label{eq: def vector valued order parameter}
\end{equation}
while its dependence on the canonical momentum
\begin{equation}
\hat{\boldsymbol{p}}\equiv
-
\mathrm{i}
\boldsymbol{\partial}\equiv
-
\mathrm{i}
\frac{
\partial
     }
     {
\partial\boldsymbol{x}
     }
\end{equation}
is explicit. 
(We have chosen natural units, i.e., $\hbar=1$.) 

The Hermiticity of the single-particle Hamiltonian
$\mathcal{H}
\big(
\hat{\boldsymbol{p}},
\boldsymbol{\varphi}(\boldsymbol{x})
\big)$ 
implies the existence of a conserved quasiparticle charge, 
which we construct explicitly with the help of
the second quantized Hamiltonian%
~\cite{footnote: comment of BCS second quantized} 
\begin{subequations}
\label{eq: def second quantized H}
\begin{equation}
\widehat{H}\equiv
\int\limits_{\boldsymbol{x}}
\widehat{\Psi}^{\dag}(\boldsymbol{x})\,
\mathcal{H}
\big(
\hat{\boldsymbol{p}},
 \boldsymbol{\varphi}(\boldsymbol{x})
\big)\,
\widehat{\Psi}      (\boldsymbol{x}).
\label{eq: def second quantized H a}
\end{equation}
The creation operators 
$\widehat{\Psi}^{\dag}(\boldsymbol{x})$
and the annihilation operators
$\widehat{\Psi}      (\boldsymbol{x})$
obey here the fermion algebra
\begin{equation}
\begin{split}
&
\left\{
\widehat{\Psi}^{\   }_{r }(\boldsymbol{x}),
\widehat{\Psi}^{\dag}_{r'}(\boldsymbol{x}')
\right\}=
\delta^{\,}_{r,r'}
\delta(\boldsymbol{x}-\boldsymbol{x}'),
\\
&
\left\{
\widehat{\Psi}^{\   }_{r }(\boldsymbol{x}),
\widehat{\Psi}^{\   }_{r'}(\boldsymbol{x}')
\right\}=
0,
\qquad r,r'=1,\cdots,R,
\end{split}
\label{eq: def second quantized H b}
\end{equation}
in some $R$-dimensional representation of the 
charge-conjugation-symmetric single-particle
Hamiltonian 
\begin{equation}
\begin{split}
\mathcal{H}
\big(
\hat{\boldsymbol{p}},
 \boldsymbol{\varphi}(\boldsymbol{x})
\big)
=&\,
-\,
\mathcal{C}^{-1}\,
\mathcal{H}
\big(
\hat{\boldsymbol{p}},
\boldsymbol{\varphi}(\boldsymbol{x})
\big)\,
\mathcal{C}.
\end{split}
\label{eq: def second quantized H c}
\end{equation}
\end{subequations}
The norm-preserving operation for charge conjugation in the
$R$-dimensional single-particle representation is denoted
by $\mathcal{C}$. The short-hand notation 
$\int\limits_{\boldsymbol{x}}$ 
stands for the space integration $\int\mathrm{d}^{d}\boldsymbol{x}$.
The definition of the conserved quasiparticle charge requires
filling of the Fermi sea for the quasiparticles created by
the field $\widehat{\Psi}^{\dag}_{r}(\boldsymbol{x})$.

We begin with an expression for a 
quasiparticle charge density in the vicinity of a point $\boldsymbol{x}$. 
To do so, we shall assume that the static vector-valued order parameter can be
decomposed additively, i.e.,%
~\cite{footnote: comment on spin-wave app}
\begin{equation}
\boldsymbol{\varphi}(\boldsymbol{x})=
\boldsymbol{\varphi}^{\,}_{0}
+
\delta\boldsymbol{\varphi}(\boldsymbol{x}),
\label{eq: varphi = varphi0 + delta varphi}
\end{equation}
in such a way that: 
(i) the single-particle Hamiltonian 
\begin{equation}
\mathcal{H}^{\,}_{0}(\hat{\boldsymbol{p}})\equiv
\mathcal{H}
\big(
\hat{\boldsymbol{p}},
\boldsymbol{\varphi}^{\,}_{0}
\big)
\end{equation}
is translation invariant,
(ii) the single-particle eigenvalue spectrum of
$\mathcal{H}^{\,}_{0}(\hat{\boldsymbol{p}})$
is fully gaped with the gap $2\Delta^{\,}_{0}>0$, 
(iii) the changes between
$\delta\boldsymbol{\varphi}(\boldsymbol{x})$
and
$\delta\boldsymbol{\varphi}(\boldsymbol{y})$
when $\boldsymbol{x}$ and $\boldsymbol{y}$ 
are a distance of order $1/\Delta^{\,}_{0}$
apart are small.
Condition (iii) implies that the gradient of
$\delta\boldsymbol{\varphi}(\boldsymbol{x})$
can be viewed as a smooth and small perturbation 
that can be treated perturbatively in 
the gradient expansion that will follow shortly.

Condition (i) implies that
\begin{subequations}
\begin{equation}
\begin{split}
\widehat{H}^{\,}_{0}\equiv&\,
\int\limits_{\boldsymbol{x}}
\widehat{\Psi}^{\dag}(\boldsymbol{x})\,
\mathcal{H}^{\,}_{0}(\hat{\boldsymbol{p}})\,
\widehat{\Psi}      (\boldsymbol{x})
\\
=&\,
\int\limits_{\boldsymbol{p}}
\widehat{\Psi}^{\dag}(\boldsymbol{p})\,
\mathcal{H}^{\,}_{0}(\boldsymbol{p})\,
\widehat{\Psi}      (\boldsymbol{p})
\end{split}
\end{equation}
with the symmetric Fourier conventions
\begin{equation}
\begin{split}
&
\widehat{\Psi}^{\dag}(\boldsymbol{x}):=
(2\pi)^{+d/2}
\int\limits_{\boldsymbol{p}}
e^{-\mathrm{i}\boldsymbol{p}\cdot\boldsymbol{x}}\,
\widehat{\Psi}^{\dag}(\boldsymbol{p}),
\\
&
\widehat{\Psi}^{\dag}(\boldsymbol{p}):=
(2\pi)^{-d/2}
\int\limits_{\boldsymbol{x}}
e^{+\mathrm{i}\boldsymbol{p}\cdot\boldsymbol{x}}\,
\widehat{\Psi}^{\dag}(\boldsymbol{x}),
\end{split}
\end{equation}
for the annihilation (creation) operators and the asymmetric convention
\begin{equation}
\begin{split}
&
\mathcal{K}(\boldsymbol{x}):=
\int\limits_{\boldsymbol{p}}
e^{+\mathrm{i}\boldsymbol{p}\cdot\boldsymbol{x}}
\mathcal{K}(\boldsymbol{p}),
\\
&
\mathcal{K}(\boldsymbol{p}):= 
\int\limits_{\boldsymbol{x}}
e^{-\mathrm{i}\boldsymbol{p}\cdot\boldsymbol{x}}
\mathcal{K}(\boldsymbol{x}),
\end{split}
\end{equation}
\end{subequations}
for any kernel $\mathcal{K}$ 
(such as the Hamiltonian $\mathcal{H}$).
Here, $\int\limits_{\boldsymbol{p}}$ is a short-hand notation for
the momentum-space integration 
$\int\mathrm{d}^{d}\boldsymbol{p}/(2\pi)^{d}$.

Condition (ii) implies the existence of the 
characteristic length scale
$1/\Delta^{\,}_{0}$. 

Define the  quasiparticle charge density
\begin{subequations}
\label{eq: def normal ordered charge}
\begin{equation}
\rho^{\,}_{\gamma}(\boldsymbol{x}):=
\int\limits_{\gamma}\frac{\mathrm{d}\omega}{2\pi}
\left\langle \boldsymbol{x}\left|
\mathrm{tr}^{\,}_{R}
\left[
\mathcal{G}         (\omega)
-
\mathcal{G}^{\,}_{0}(\omega)
\right]
\right|\boldsymbol{x}\right\rangle.
\label{eq: def normal ordered charge a}
\end{equation}
Here, $\mathrm{tr}^{\,}_{R}$ denotes the trace over the
$R$-dimensional degrees of freedom and
we have introduced the Euclidean single-particle Green functions
\begin{equation}
\mathcal{G}         (\omega):=
\frac{
1
     }
     {
\mathrm{i}\omega - \mathcal{H}
     },
\qquad
\mathcal{G}^{\,}_{0}(\omega):=
\frac{
1
     }
     {
\mathrm{i}\omega - \mathcal{H}^{\,}_{0}
     }.
\label{eq: def normal ordered charge b}
\end{equation}
\end{subequations}
The  quasiparticle charge density 
$\rho^{\,}_{\gamma}(\boldsymbol{x})$ 
depends on the contour of integration
$\gamma$. The latter is chosen as in
Appendix~\ref{appsec: Schrieffer's counting argument}
so that the integration over 
the $\omega$-complex plane picks up 
\textit{only the first-order poles from the
non-vanishing and negative energy eigenvalues} of $\mathcal{H}$
and of $\mathcal{H}^{\,}_{0}$
[see Eq.~(\ref{eq:application of the residue them given gamma})].

It is shown in Appendix~\ref{appsec: Schrieffer's counting argument}
[see Eq.~(\ref{eq: global sum rule with Green c})]
that
Eq.~(\ref{eq: def second quantized H c})
and condition (ii) imply that
the \textit{total number $N$ of unoccupied zero modes} 
of $\mathcal{H}$ is related to the 
conserved quasiparticle charge 
$Q^{\,}_{\gamma}$ by
\begin{equation}
\frac{N}{2}=
-
\int\mathrm{d}^{d}\boldsymbol{x}\,
\rho^{\,}_{\gamma}(\boldsymbol{x})\equiv
-
Q^{\,}_{\gamma}.
\label{eq: master counting formula}
\end{equation}
According to the counting formula%
~(\ref{eq: master counting formula}),
computing $N$ reduces to computing the 
conserved quasiparticle charge 
$Q^{\,}_{\gamma}$ 
induced by the smooth variation of
$\delta\boldsymbol{\varphi}(\boldsymbol{x})$
through space defined in Eq.~(\ref{eq: def space}).
The Hermiticity
of the single-particle Hamiltonian%
~(\ref{eq: def second quantized H c})
guarantees the existence of this 
conserved quasiparticle charge. 
A local law for the conservation of quasiparticle current 
follows when the single-particle Hamiltonian is local.
For single-particle Hamiltonians with the global U(1)
symmetry delivering the conservation of the total electric
charge, the conserved quasiparticle charge 
$Q^{\,}_{\gamma}$ 
is nothing but the electric charge 
in units in which the electron charge is unity. 
For single-particle Hamiltonians describing
Bogoliubov-de-Gennes (BdG) quasiparticles, this global U(1)
symmetry is spontaneously broken. The 
 conserved quasiparticle charge $Q^{\,}_{\gamma}$ 
is then related to the conserved thermal current
of BdG quasiparticles. The counting formula%
~(\ref{eq: master counting formula})
appears implicitly in 
Ref.~\onlinecite{Su81} and
explicitly in Ref.~\onlinecite{Jackiw81a}.%
~\cite{Wilczek02} 

There is an alternative to specifying $\gamma$ 
in the counting formula%
~(\ref{eq: master counting formula}).
We can regulate the first-order pole of the Green
function $\mathcal{G}(\omega)$ at $\omega=0$
by adding a perturbation that moves all zero modes
to strictly positive energies. This perturbation must be small
if all these positive energy increments are to
remain much smaller than the threshold $\Delta^{\,}_{0}$
to the continuum. We can then safely replace the contour of
integration $\gamma$ in the counting formula%
~(\ref{eq: master counting formula})
by $\mathbb{R}$ 
since the subtraction of $\mathcal{G}^{\,}_{0}$ 
from $\mathcal{G}$ insures the
convergence of the $\omega$ integration for large $\omega$.

For example, we imagine that it is possible to
augment the vector-valued order parameter%
~(\ref{eq: varphi = varphi0 + delta varphi})
by the conjugation-symmetry-breaking real-valued field $\phi$
without loosing conditions (i)-(iii).
More precisely, we define the $(D+1)$-tuplet
\begin{subequations}
\label{eq: def varphi to phi}
\begin{equation}
\boldsymbol{\phi}(\boldsymbol{x})\equiv
\begin{pmatrix}
\phi^{\,}_{1}(\boldsymbol{x})\\
\vdots\\
\phi^{\,}_{D}(\boldsymbol{x})\\
\phi^{\,}_{D+1}\\
\end{pmatrix}
\equiv
\begin{pmatrix}
\varphi^{\,}_{1}(\boldsymbol{x})\\
\vdots\\
\varphi^{\,}_{D}(\boldsymbol{x})\\
\phi\\
\end{pmatrix}
\label{eq: def varphi to phi a}
\end{equation}
and we assume that Eq.~(\ref{eq: def second quantized H c}) 
becomes
\begin{equation}
\mathcal{H}
\big(
\hat{\boldsymbol{p}},
\boldsymbol{\varphi}(\boldsymbol{x}),
\phi
\big)
=
-\,
\mathcal{C}^{-1}\,
\mathcal{H}
\big(
\hat{\boldsymbol{p}},
\boldsymbol{\varphi}(\boldsymbol{x}),
-\phi
\big)\,
\mathcal{C}.
\label{eq: def second quantized H c bis}
\end{equation}
\end{subequations}
It then follows that
\begin{equation}
Q(\phi):=
\int\mathrm{d}^{d}\boldsymbol{x}\,
\rho(\boldsymbol{x},\phi)=
-Q(-\phi)
\end{equation}
where the quasiparticle charge density
$\rho(\boldsymbol{x},\phi)$ is obtained from 
Eq.~(\ref{eq: def normal ordered charge}) with
$\mathbb{R}$ substituting for $\gamma$,
Hamiltonian%
~(\ref{eq: def second quantized H c bis})
substituting for
$
\mathcal{H}
\big(
\hat{\boldsymbol{p}},
\boldsymbol{\varphi}(\boldsymbol{x})
\big)
$,
and
$
\mathcal{H}^{\,}_{0}\equiv
\mathcal{H}
\big(
\hat{\boldsymbol{p}},
\boldsymbol{\varphi}^{\,}_{0},
\phi
\big)
$
substituting for
$
\mathcal{H}^{\,}_{0}\equiv
\mathcal{H}
\big(
\hat{\boldsymbol{p}},
\boldsymbol{\varphi}^{\,}_{0}
\big)
$.
The smoking gun for the \textit{unoccupied zero modes} 
is now a discontinuity at the conjugation-symmetric point 
$\phi=0$ of the odd function $Q(\phi)$ of $\phi$, 
i.e., the counting formula%
~(\ref{eq: master counting formula}) 
has become
\begin{equation}
\frac{N}{2}=
-
\lim_{\phi\to0}
\int\mathrm{d}^{d}\boldsymbol{x}\,
\rho(\boldsymbol{x},\phi)\equiv
-
\lim_{\phi\to0}
Q(\phi)
\label{eq: master counting formula bis}
\end{equation}
where the sign of $\phi$ is to be chosen so as to 
move the $N$ zero modes along the energy axis 
to positive energies.
Furthermore, we can relax the condition that the
symmetry breaking $\phi$ is constant everywhere
in space ($\boldsymbol{x}$) provided that the condition
\begin{equation}
\boldsymbol{\phi}^{2}(\boldsymbol{x})\approx
\boldsymbol{\varphi}^{2}_{0}\equiv
\Delta^{2}_{0}
\end{equation}
holds everywhere in $\mathbb{R}^{d}$.
Condition (iii) then means that 
$\delta\boldsymbol{\phi}(\boldsymbol{x})$
varies slowly on the  characteristic length scale
$1/\Delta^{\,}_{0}$. 

Rather than computing $\rho(\boldsymbol{x})$ exactly,
say with the help of numerical tools,
we are after the leading contribution 
to the gradient expansion of the 
quasiparticle charge density $\rho(\boldsymbol{x})$,
which we shall denote as
$\rho^{\,}_{\mathrm{adia}}(\boldsymbol{x})$
where the subscript ``adia''
refers to the adiabatic approximation
contained in condition (iii).

The order parameter~(\ref{eq: def varphi to phi a})
enters linearly in all the single-particle Hamiltonians
that we shall consider explicitly in this paper. Hence, 
there follows the additive law
\begin{equation}
\begin{split}
\mathcal{H}
\big(
\hat{\boldsymbol{p}},
\boldsymbol{\phi}(\boldsymbol{x})
\big)
=&\,
\mathcal{H}
\big(
\hat{\boldsymbol{p}},
\boldsymbol{\phi}^{\,}_{0}
\big)
+
\mathcal{V}
\big(
\hat{\boldsymbol{p}},
\delta\boldsymbol{\phi}(\boldsymbol{x})
\big)
\\
\equiv&\,
\mathcal{H}^{\,}_{0}
+
\mathcal{V}
\big(
\hat{\boldsymbol{p}},
\delta\boldsymbol{\phi}(\boldsymbol{x})
\big)
\end{split}
\end{equation}
upon insertion of
\begin{equation}
\boldsymbol{\phi}(\boldsymbol{x})=
\boldsymbol{\phi}^{\,}_{0}
+
\delta\boldsymbol{\phi}(\boldsymbol{x}).
\label{eq: additive decomposition of phi}
\end{equation}
Upon second-quantization, this implies that
\begin{subequations}
\begin{equation}
\begin{split}
\widehat{H}=&\,
\int\limits_{\boldsymbol{p}}
\int\limits_{\boldsymbol{q}}
\widehat{\Psi}^{\dag}(\boldsymbol{p})
\left[
\mathcal{H}^{\,}_{0}(\boldsymbol{p})\,
\delta(\boldsymbol{p}-\boldsymbol{q})
+
\mathcal{V}         (\boldsymbol{p};\boldsymbol{q})
\right]
\widehat{\Psi}      (\boldsymbol{q})
\end{split}
\end{equation}
holds with
\begin{equation}
\mathcal{V}(\boldsymbol{p};\boldsymbol{q}):=
\left(
\frac{
\partial\mathcal{H}
     }
     {
\partial\boldsymbol{\phi}
     }
\right)^{\,}_{0}
\left(
\frac{
\boldsymbol{p}+\boldsymbol{q}
     }
     {
2
     }
\right)
\cdot
\delta
\boldsymbol{\phi}(\boldsymbol{p}-\boldsymbol{q}).
\label{eq: def mathcal V}
\end{equation}
The subscript $0$ 
means setting $\delta\boldsymbol{\phi}$ to zero
so that the gradient
\begin{equation}
\left( 
\frac{
\partial\mathcal{H}
     }
     {
\partial\boldsymbol{\phi}
     }
\right)^{\,}_{0}
=
-
\left(
\frac{
\partial\mathcal{G}^{-1}
     }
     {
\partial\boldsymbol{\phi}
     }
\right)^{\,}_{0}
\end{equation}
depends only on the single-particle canonical momentum operator
$\hat{\boldsymbol{p}}\equiv-\mathrm{i}\boldsymbol{\partial}$.
We have also adopted the convention that matrix elements of
$\hat{\boldsymbol{p}}$ are to be symmetrized, i.e.,
\begin{equation}
\left(
f^{*}
\hat{\boldsymbol{p}}
g    
\right)
(\boldsymbol{x})\equiv
-
\frac{\mathrm{i}}{2}
\Bigg(
f^{*}
\left(
\frac{\partial g}{\partial\boldsymbol{x}}
\right)
-
\left(
\frac{\partial f^{*}}{\partial\boldsymbol{x}}
\right)
g    
\Bigg)
(\boldsymbol{x})
\end{equation}
\end{subequations}
for any differentiable functions $f$ and $g$.

We now expand the quasiparticle charge density
up to the first non-trivial order in an expansion in powers of
$\mathcal{V}$ with the help of the geometrical series
\begin{equation}
\mathcal{G}(\omega)
-
\mathcal{G}^{\,}_{0}(\omega)\approx
\sum_{n=1}^{\infty}
\left(
\mathcal{G}^{\,}_{0}(\omega)
\mathcal{V}
\right)^{n}
\mathcal{G}^{\,}_{0}(\omega).
\label{eq: geometric expansion Green function diff}
\end{equation}
This gives the expansion
\begin{subequations}
\label{eq: master formula before expanding in momenta}
\begin{equation}
\rho(\boldsymbol{x})\approx
\sum_{n=1}^{\infty}
\rho^{\,}_{n}(\boldsymbol{x})
\label{eq: master formula before expanding in momenta a}
\end{equation}
with
\begin{equation}
\begin{split}
\rho^{\,}_{n}(\boldsymbol{x}):=&\,
\int\limits_{\omega}
\int\limits_{\boldsymbol{p}}
\int\limits_{\boldsymbol{q}^{\,}_{1}}
\cdots
\int\limits_{\boldsymbol{q}^{\,}_{n}}\,
e^{
\mathrm{i}
\left(
\boldsymbol{q}^{\,}_{1}
+
\cdots
+
\boldsymbol{q}^{\,}_{n}
\right)
\cdot
\boldsymbol{x}
  }
\\
&
\times
\sum_{\mathsf{a}^{\,}_{1},\cdots,\mathsf{a}^{\,}_{n}=1}^{D+1}
I^{\,}_{\mathsf{a}^{\,}_{n},\cdots,\mathsf{a}^{\,}_{1}}
(\omega,\boldsymbol{p},\boldsymbol{q}^{\,}_{1},\cdots,
\boldsymbol{q}^{\,}_{n})
\\
&
\times\,
\delta\phi^{\,}_{\mathsf{a}^{\,}_{n}}(\boldsymbol{q}^{\,}_{n})
\cdots
\delta\phi^{\,}_{\mathsf{a}^{\,}_{1}}(\boldsymbol{q}^{\,}_{1})
\end{split}
\label{eq: master formula before expanding in momenta b}
\end{equation}
where $\int\limits_{\omega}\equiv\int\frac{\mathrm{d}\omega}{2\pi}$
and the integrand   
\begin{equation}
\begin{split}
&
I^{\,}_{\mathsf{a}^{\,}_{n},\cdots,\mathsf{a}^{\,}_{1}}
(\omega,\boldsymbol{p},\boldsymbol{q}^{\,}_{1},\cdots,\boldsymbol{q}^{\,}_{n})
\\
&\hphantom{A}:=
\mathrm{tr}^{\,}_{R}
\Bigg[
\mathcal{G}^{\,}_{0}
\left(
\omega,
\boldsymbol{p}
+
\boldsymbol{q}^{\,}_{1}
+
\boldsymbol{q}^{\,}_{2}
+
\cdots
+
\boldsymbol{q}^{\,}_{n}
\right) 
\\
&\hphantom{AA}\times
\left(
\frac{
\partial\mathcal{H}
     }
     {
\partial\phi^{\,}_{\mathsf{a}^{\,}_{n}}
     }
\right)^{\,}_{0}
\left(
\omega,
\boldsymbol{p}
+
\boldsymbol{q}^{\,}_{1}
+
\boldsymbol{q}^{\,}_{2}
+
\cdots
+
\boldsymbol{q}^{\,}_{n-1}
+
\frac{
\boldsymbol{q}^{\,}_{n}
     }
     {
2
     }
\right)
\\ 
&\hphantom{AA}\times
\mathcal{G}^{\,}_{0}
\left(
\omega,
\boldsymbol{p}
+
\boldsymbol{q}^{\,}_{1}
+
\boldsymbol{q}^{\,}_{2}
+
\cdots
+
\boldsymbol{q}^{\,}_{n-1}
\right)
\\
&\hphantom{AA}\times
\left(
\frac{
\partial\mathcal{H}
     }
     {
\partial\phi^{\,}_{\mathsf{a}^{\,}_{n-1}}
     }
\right)^{\,}_{0}
\left(
\omega,
\boldsymbol{p}
+
\boldsymbol{q}^{\,}_{1}
+
\boldsymbol{q}^{\,}_{2}
+
\cdots
+
\boldsymbol{q}^{\,}_{n-2}
+
\frac{
\boldsymbol{q}^{\,}_{n-1}
     }
     {
2
     }
\right)
\\
&\hphantom{AAA}
\vdots
\\
&\hphantom{AA}\times
\mathcal{G}^{\,}_{0}
\left(
\omega,
\boldsymbol{p}
+
\boldsymbol{q}^{\,}_{1}
\right)
\left(
\frac{
\partial\mathcal{H}
     }
     {
\partial\phi^{\,}_{\mathsf{a}^{\,}_{1}}
     }
\right)^{\,}_{0}
\left(
\omega,
\boldsymbol{p}
+
\frac{
\boldsymbol{q}^{\,}_{1}
     }
     {
2
     }
\right)
\mathcal{G}^{\,}_{0}
\left(
\omega,
\boldsymbol{p}
\right)
\Bigg].
\end{split}
\label{eq: master formula before expanding in momenta c}
\end{equation}
\end{subequations}

Finally, we expand the integrand%
~(\ref{eq: master formula before expanding in momenta c})
in powers of the coordinates 
$q^{\,}_{1i^{\,}_{1}}$,
$\cdots$,
$q^{\,}_{ni^{\,}_{n}}$
of the momenta 
$\boldsymbol{q}^{\,}_{1}$,
$\cdots$,
$\boldsymbol{q}^{\,}_{n}$,
which are then to be integrated over. 
Condition (iii) suggests that we keep only first-order
derivatives in the slowly varying fluctuations 
$\delta\boldsymbol{\phi}(\boldsymbol{x})$
of the order parameter~(\ref{eq: varphi = varphi0 + delta varphi}).
This approximation yields the leading contribution 
$\rho^{\,}_{\mathrm{adia}}(\boldsymbol{x})$
in the gradient expansion of the 
quasiparticle charge density%
~(\ref{eq: master formula before expanding in momenta a})
and becomes exact in the limit when the ratio between
the characteristic length scale $1/\Delta^{\,}_{0}$ 
and the characteristic length scale over which the change in
$\delta\boldsymbol{\phi}(\boldsymbol{x})$
becomes significant vanishes. 

However, to each order $n$ in this expansion there are terms
for which not all $\delta\boldsymbol{\phi}(\boldsymbol{x})$ 
are differentiated. These terms do not have to be small.
Hence, they should be treated non-perturbatively.
This is achieved by replacing
the Euclidean single-particle Green function
\begin{equation}
\mathcal{G}^{\,}_{0}(\omega,\boldsymbol{p}):=
\frac{
1
     }
     {
\mathrm{i}\omega
-
\mathcal{H}\big(\boldsymbol{p},\boldsymbol{\phi}^{\,}_{0}\big)
     }
\label{eq: def mathcal{G}{0}}
\end{equation}
with the Euclidean semi-classical Green function
\begin{equation}
\mathcal{G}^{\,}_{\text{s-c}}(\omega,\boldsymbol{p},\boldsymbol{x}):=
\frac{
1
     }
{
\mathrm{i}\omega
-
\mathcal{H}
\big(
\boldsymbol{p},
\boldsymbol{\phi}(\boldsymbol{x})
\big)
     }
\label{eq: def semi classical G}
\end{equation}
to any given order $n$ in the expansion%
~(\ref{eq: master formula before expanding in momenta}).
This result is rooted in the fact that the choice
of $\boldsymbol{\phi}^{\,}_{0}$ in the additive decomposition%
~(\ref{eq: additive decomposition of phi}) 
is arbitrary from the point of view of the expansion%
~(\ref{eq: master formula before expanding in momenta}).
Such arbitrariness should not appear in the final result,
i.e., the final result should only depend on 
$\boldsymbol{\phi}(\boldsymbol{x})$ 
or
$\partial^{\,}_{i}\phi^{\ }_{\mathsf{a}}(\boldsymbol{x})\equiv
\frac{\partial\phi^{\ }_{\mathsf{a}}}{\partial x^{i}}
(\boldsymbol{x})$ with $\mathsf{a}=1,\cdots,d+1$ and
$i=1,\cdots,d$.%
\cite{footnote: semi-classical gradient expansion}

Thus, by collecting the appropriate derivatives of the spatially varying
order parameter $\delta\boldsymbol{\phi}(\boldsymbol{x})$, 
we arrive at expressions for the induced
quasiparticle charge density. 
We now analyze this expression according to which term is 
the first non-vanishing contribution to the expansion. 

When the adiabatic approximation for the 
quasiparticle charge density is the non-vanishing $n=1$ term in the
gradient expansion, it is given by
\begin{subequations}
\label{eq: n=1 adia appro}
\begin{equation}
\rho^{\,}_{\mathrm{adia}}(\boldsymbol{x})=
\mathcal{I}^{\,}_{i^{\,}_{1}\mathsf{a}^{\,}_{1}}
(-\mathrm{i})
\left(
\partial^{\,}_{i^{\,}_{1}}\phi^{\,}_{\mathsf{a}^{\,}_{1}}
\right)(\boldsymbol{x}).
\label{eq: n=1 adia appro b}
\end{equation}
Here, the summation convention is assumed over repeated 
indices and the coefficients are given by
\begin{equation}
\mathcal{I}^{\,}_{i^{\,}_{1}\mathsf{a}^{\,}_{1}}:=
\int\limits_{\omega}
\int\limits_{\boldsymbol{p}}
\frac{1}{2}
\mathrm{tr}^{\,}_{R}
\left(
\left[
\frac{
\partial\mathcal{G}^{-1}
     }
     {
\partial \phi^{\,}_{\mathsf{a}^{\,}_{1}}
     },
\frac{
\partial\mathcal{G}
     }
     {
\partial p^{\,}_{i^{\,}_{1}}
     }
\right]
\mathcal{G}
\right)^{\,}_{0}
(\omega,\boldsymbol{p}).
\label{eq: n=1 adia appro a}
\end{equation}
\end{subequations}
The subscript 0 refers to the semi-classical
Green function~(\ref{eq: def semi classical G}).
This case is the relevant one for the study of point defects in
one-dimensional space.

When the first non-vanishing contribution to the adiabatic expansion
occurs at $n=2$, then the  quasiparticle charge
density contains two gradients and is given by
\begin{subequations}
\label{eq: n=2 adia appro}
\begin{equation}
\rho^{\,}_{\mathrm{adia}}(\boldsymbol{x})=
\mathcal{I}^{\,}_{i^{\,}_{2}\mathsf{a}^{\,}_{2}i^{\,}_{1}\mathsf{a}^{\,}_{1}}
(-\mathrm{i})^{2}
\left(
\partial^{\,}_{i^{\,}_{2}}\phi^{\,}_{\mathsf{a}^{\,}_{2}}
\right)
\left(
\partial^{\,}_{i^{\,}_{1}}\phi^{\,}_{\mathsf{a}^{\,}_{1}}
\right)(\boldsymbol{x}).
\label{eq: n=2 adia appro b}
\end{equation}
By assumption,
$\mathcal{I}^{\,}_{i^{\,}_{1}\mathsf{a}^{\,}_{1}}$
defined by Eq.~(\ref{eq: n=1 adia appro a}) 
vanishes, but 
\begin{equation}
\begin{split}
\mathcal{I}^{\,}_{i^{\,}_{2}\mathsf{a}^{\,}_{2}i^{\,}_{1}\mathsf{a}^{\,}_{1}}:=&\,
-
\int\limits_{\omega}
\int\limits_{\boldsymbol{p}}
\frac{1}{2}
\mathrm{tr}^{\,}_{R}
\Bigg(
2
\frac{
\partial\mathcal{G}
     }
     {
\partial p^{\,}_{i^{\,}_{2}}
     }
\frac{
\partial\mathcal{G}^{-1}
     }
     {
\partial \phi^{\,}_{\mathsf{a}^{\,}_{2}}
     }
\mathcal{G}
\frac{
\partial\mathcal{G}^{-1}
     }
     {
\partial \phi^{\,}_{\mathsf{a}^{\,}_{1}}
     }
\frac{
\partial\mathcal{G}
     }
     {
\partial p^{\,}_{i^{\,}_{1}}
     }
\\
&\,
+
\mathcal{G}
\frac{
\partial^{2}\mathcal{G}^{-1}
     }
     {
\partial p^{\,}_{i^{\,}_{2}}
\partial \phi^{\,}_{\mathsf{a}^{\,}_{2}}
     }
\mathcal{G}
\frac{
\partial\mathcal{G}^{-1}
     }
     {
\partial \phi^{\,}_{\mathsf{a}^{\,}_{1}}
     }
\frac{
\partial\mathcal{G}
     }
     {
\partial p^{\,}_{i^{\,}_{1}}
     }
\\
&\,
+
\frac{
\partial\mathcal{G}
     }
     {
\partial p^{\,}_{i^{\,}_{2}}
     }
\frac{
\partial\mathcal{G}^{-1}
     }
     {
\partial \phi^{\,}_{\mathsf{a}^{\,}_{2}}
     }
\mathcal{G}
\frac{
\partial^{2}\mathcal{G}^{-1}
     }
     {
\partial p^{\,}_{i^{\,}_{1}}
\partial \phi^{\,}_{\mathsf{a}^{\,}_{1}}
     }
\mathcal{G}
\Bigg)^{\,}_{0}
(\omega,\boldsymbol{p})
\end{split}
\label{eq: n=2 adia appro a}
\end{equation}
\end{subequations}
does not. Again, the summation convention is assumed over repeated
indices and the subscript 0 refers to the semi-classical
Green function~(\ref{eq: def semi classical G}).
We have chosen to represent
Eq.~(\ref{eq: n=2 adia appro})
so as to make the reality of
$\rho^{\,}_{\mathrm{adia}}(\boldsymbol{x})$ 
manifest. This case is the relevant one for the study of point defects in
two-dimensional space. 

Observe that, whenever $n>1$, 
we must allow for the possibility that
\begin{equation}
\frac{
\partial^{2}\mathcal{H}
     }
     {
\partial p^{\,}_{i}
\partial \phi^{\,}_{\mathsf{a}}
     }
\equiv
-
\frac{
\partial^{2}\mathcal{G}^{-1}
     }
     {
\partial p^{\,}_{i}
\partial \phi^{\,}_{\mathsf{a}}
     }
\label{eq: non-relativistic term}
\end{equation}
is non-vanishing for some $i=1,\cdots,d$ and some 
$\mathsf{a}=1,\cdots,D+1$.
These terms occur when dealing with a $\mathsf{p}$-wave superconductor
in $d=2$ dimensions as we do in Sec.%
~\ref{sec: Chiral mathsf{p}-wave superconductor when d=2}; however, we
find by explicit calculation that these terms vanish upon integration
over $\omega$ and $\boldsymbol{p}$.

When the order parameter is independent of momentum, 
\begin{subequations}
\label{eq: leading n=d order if no double derivative}
\begin{equation}
\frac{
\partial^{2}\mathcal{H}
     }
     {
\partial p^{\,}_{i}
\partial \phi^{\,}_{\mathsf{a}}
     }
\equiv
-
\frac{
\partial^{2}\mathcal{G}^{-1}
     }
     {
\partial p^{\,}_{i}
\partial \phi^{\,}_{\mathsf{a}}
     }
=0
\end{equation}
for any $i=1,\cdots,d$ and any $\mathsf{a}=1,\cdots,D+1$.
For example this is the case with $D=d$
for Dirac fermions in $d$-dimensional 
space interacting with $(d+1)$ real-valued Higgs fields, 
in which case
it is the coefficient%
~\cite{footnote: gradient expansion for Dirac}
\begin{equation}
\begin{split}
\mathcal{I}^{\;}_{
i^{\,}_{n}\mathsf{a}^{\,}_{n}\cdots i^{\,}_{1}\mathsf{a}^{\,}_{1}
                }=&-\mathrm{i}
\int\limits_{\omega}
\int\limits_{\boldsymbol{p}}
\mathrm{tr}^{\,}_{R}
\Bigg[
\left(
\mathcal{G}\frac{\partial\mathcal{G}^{-1}}{\partial p^{\,}_{i^{\,}_{n}}}
\mathcal{G}\frac{\partial\mathcal{G}^{-1}}{\partial \phi^{\,}_{\mathsf{a}^{\,}_{n}}}
\right)
\cdots
\\
&
\cdots
\left(
\mathcal{G}\frac{\partial\mathcal{G}^{-1}}{\partial p^{\,}_{i^{\,}_{1}}}
\mathcal{G}\frac{\partial\mathcal{G}^{-1}}{\partial \phi^{\,}_{\mathsf{a}^{\,}_{1}}}
\right)
\left(
\mathcal{G}\frac{\partial\mathcal{G}^{-1}}{\partial\omega}
\right)
\Bigg]^{\,}_{0}(\omega,\boldsymbol{p})
\end{split}
\end{equation}
that controls the adiabatic approximation to $n$-th order 
through 
\begin{equation}
\rho^{\,}_{\mathrm{adia}}(\boldsymbol{x})=
(-\mathrm{i})^{d}
\mathcal{I}^{\;}_{
i^{\,}_{n}\mathsf{a}^{\,}_{n}\cdots i^{\,}_{1}\mathsf{a}^{\,}_{1}
                }
\left(\partial^{\,}_{i^{\,}_{n}}\phi^{\,}_{\mathsf{a}^{\,}_{n}}\right)
\cdots
\left(\partial^{\,}_{i^{\,}_{1}}\phi^{\,}_{\mathsf{a}^{\,}_{1}}\right)
(\boldsymbol{x})
\end{equation}
\end{subequations}
as we shall show in Sec.%
~\ref{sec: Dirac single-particle mathcal{H} for any d}.
The subscript 0 refers to the semi-classical
Green function~(\ref{eq: def semi classical G}).

Finally, with the expression for the \textit{local} 
quasiparticle charge density
$\rho^{\,}_{\mathrm{adia}}(\boldsymbol{x})$ within the adiabatic
approximation in hand, we can compute the \textit{total} 
quasiparticle charge
$Q^{\,}_{\mathrm{adia}}$. Naturally, one goes from the local density to the
total charge by integrating the former over all space. 
We conclude that
\begin{subequations}
\label{eq: def semi classical Q}
\begin{equation}
\begin{split}
Q^{\,}_{\mathrm{adia}}=&\,
\int \mathrm{d}^{d}x\,
\rho^{\,}_{\mathrm{adia}}(\boldsymbol{x})
\\
=&\,
\int\mathrm{d}^{d}x\,
(-\mathrm{i})^{d}
\;
\mathcal{I}^{\;}_{i^{\,}_{n}\mathsf{a}^{\,}_{n}
\cdots i^{\,}_{1}\mathsf{a}^{\,}_{1}}
(\boldsymbol{x})
\\
&\times
\left(\partial^{\,}_{i^{\,}_{n}}\phi^{\,}_{\mathsf{a}^{\,}_{n}}\right)
\cdots
\left(\partial^{\,}_{i^{\,}_{1}}\phi^{\,}_{\mathsf{a}^{\,}_{1}}\right)
(\boldsymbol{x}),
\end{split}
\label{eq: def semi classical Q a}
\end{equation}
where it is the Euclidean semi-classical Green function
\begin{equation}
\mathcal{G}^{\,}_{\text{s-c}}(\omega,\boldsymbol{p},\boldsymbol{x}):=
\frac{
1
     }
{
\mathrm{i}\omega
-
\mathcal{H}
\big(
\boldsymbol{p},
\boldsymbol{\phi}(\boldsymbol{x})
\big)
     }
\label{eq: def semi classical Q b}
\end{equation}
\end{subequations}
that enters the kernel $\mathcal{I}$.

\section{
Zero modes induced by point defects
        }
\label{sec: zero modes induced by point defects}

We are going to apply the adiabatic expansion
from Sec.%
~\ref{sec: Gradient expansion}
to the case of point defects supported by the static order parameter
\begin{subequations}
\label{eq: point defects and order parameter}
\begin{equation}
\boldsymbol{\phi}(\boldsymbol{x})\equiv
\begin{pmatrix}
\phi^{\,}_{1}(\boldsymbol{x})\\
\vdots\\
\phi^{\,}_{D}(\boldsymbol{x})\\
\phi^{\,}_{D+1}(\boldsymbol{x})\\
\end{pmatrix}
\equiv
\begin{pmatrix}
\varphi^{\,}_{1}(\boldsymbol{x})\\
\vdots\\
\varphi^{\,}_{D}(\boldsymbol{x})\\
\phi(\boldsymbol{x})\\
\end{pmatrix}
\in\mathbb{R}^{d+1}
\label{eq: point defects and order parameter a}
\end{equation}
and
\begin{equation}
\boldsymbol{\phi}^{2}(\boldsymbol{x})\approx
\boldsymbol{\varphi}^{2}_{0}\equiv
\Delta^{2}_{0}
\label{eq: point defects and order parameter b}
\end{equation}
when space is $d$-dimensional, i.e., when
\begin{equation}
\boldsymbol{x}\in\mathbb{R}^{d}.
\label{eq: point defects and order parameter c}
\end{equation}
\end{subequations}
The component 
$\phi^{\,}_{D+1}(\boldsymbol{x})\equiv\phi(\boldsymbol{x})$ 
breaks locally the charge
conjugation symmetry. This component determines if a zero mode
bound to a defect at $\boldsymbol{x}$
is occupied or unoccupied. All remaining components of the
order parameter~(\ref{eq: point defects and order parameter a})
are compatible with the charge-conjugation symmetry
of the single-particle Hamiltonian. The condition%
~(\ref{eq: point defects and order parameter b})
suggests that the homotopy group
\begin{equation}
\Pi^{\,}_{d}(S^{d})=\mathbb{Z}
\end{equation}
of smooth maps from the compactification of $\mathbb{R}^{d}$
into the $d$-sphere $S^{d}$  
to the $d$-sphere $S^{d}$ in order-parameter space $\mathbb{R}^{d+1}$
might play an important role.%
~\cite{footnote: def m sphere}

We begin in one dimensional space ($x$) 
with a generic single-particle Hamiltonian.
We show that the conserved quasiparticle charge 
$Q^{\,}_{\mathrm{adia}}$
that enters the counting formula%
~(\ref{eq: master counting formula}) 
is \textit{the first Chern number} 
if we relax the condition that
the charge-symmetry-breaking component $\phi$ 
of the order parameter is infinitesimally small
and if we compactify space%
~(\ref{eq: point defects and order parameter c})
and the order-parameter space%
~(\ref{eq: point defects and order parameter a}).

We continue with the chiral $\mathsf{p}$-wave superconductor in
two-dimensions when the superconducting order parameter supports a vortex.
We show that the number of unoccupied
zero modes bound to a vortex with unit vorticity computed from
the adiabatic approximation agrees with the direct construction 
of zero modes once all microscopic data have been supplied. 
Moreover, we show that the conserved quasiparticle charge 
$Q^{\,}_{\mathrm{adia}}$ that enters the counting formula%
~(\ref{eq: master counting formula})  
is also related to \textit{the second Chern number}
after compactification
of both space%
~(\ref{eq: point defects and order parameter c}) 
and order-parameter space%
~(\ref{eq: point defects and order parameter a}).
This is apriori 
surprising since the superconducting order parameter couples 
to the momentum contrary to the simpler case of Dirac fermions.

We also study Dirac single-particle Hamiltonians in $d$-dimensional
space%
~(\ref{eq: point defects and order parameter c})
that are represented by Dirac matrices of dimension $R=2^{d}$. 
We show how the conserved quasiparticle charge 
$Q^{\,}_{\mathrm{adia}}$ that enters the counting formula%
~(\ref{eq: master counting formula})  
is related to \textit{the $d$-th Chern number} 
when the Dirac spinor couples to
a $(d+1)$-tuplet of Higgs field.

We close by discussing how to interpret 
the adiabatic approximation.

\subsection{
Generic single-particle $\mathcal{H}$ when $d=1$
           }
\label{subsec: Generic single-particle mathcal{H} when d=1}

We compactify space and the order-parameter space,
\begin{equation}
x\in S^{1},
\qquad
\boldsymbol{\phi}(\theta)\in S^{1}\subset\mathbb{R}^{2},
\end{equation}
respectively,
such that
\begin{equation}
\boldsymbol{\phi}(x):=
\begin{pmatrix}
\phi^{\,}_{1}(x)
\\
\phi^{\,}_{2}(x)
\end{pmatrix} 
= 
m
\begin{pmatrix}
\sin\theta(x)
\\
\cos\theta(x)
\end{pmatrix},
\end{equation}
and denote with $\mathcal{G}=(\mathrm{i}\omega-\mathcal{H})^{-1}$
the single-particle Green function in Euclidean space for
any suitable $R\times R$ matrix-valued single-particle Hamiltonian 
$\mathcal{H}\big(\hat{p},\boldsymbol{\phi}(x)\big)$.
Suitability means that the dependence on the momentum operator
$\hat{p}\equiv-\mathrm{i}\partial^{\,}_{x}$
in $\mathcal{H}\big(\hat{p},\boldsymbol{\phi}(x)\big)$
is only restricted by locality while that on
the two-component order parameter
$\boldsymbol{\phi}(x)$ is linear.
Furthermore,
$\mathcal{H}\big(\hat{p},\boldsymbol{\phi}(x)\big)$
obeys Eq.~(\ref{eq: def varphi to phi}) for $D=1$
and supports the spectral gap $2m$ for a uniform
$\boldsymbol{\phi}^{\,}_{0}$.
Under these conditions, for any $p$
the semi-classical $R\times R$ matrix
$\mathcal{H}\big(p,\boldsymbol{\phi}(x)\big)$
is traceless and its square is proportional to the unit matrix
with the smallest nonvanishing eigenvalue no smaller than $m^{2}$.
We have introduced the spherical coordinate
$\theta$
of the one-sphere $S^{1}\subset\mathbb{R}^{2}$
in order-parameter space.

According to Eq.~(\ref{eq: n=1 adia appro}),
the conserved quasiparticle charge 
$Q^{\,}_{\mathrm{adia}}$ becomes
\begin{equation}
\begin{split}
Q^{\,}_{\mathrm{adia}}=&\, 
\int\limits_{\omega}
\int\limits_{p\in S^{1}}
\int\limits_{x\in S^{1}}
\frac{1}{2}
\mathrm{tr}^{\,}_{R}
\Big(
\mathcal{G}\partial^{\,}_{p}\mathcal{G}^{-1}
\mathcal{G}\partial^{\,}_{\theta}\mathcal{G}^{-1}
\mathcal{G}
\\
&\,
-
\mathcal{G}\partial^{\,}_{\theta}\mathcal{G}^{-1}
\mathcal{G}\partial^{\,}_{p}\mathcal{G}^{-1}
\mathcal{G}
\Big)^{\,}_{0}
(-\mathrm{i})
\partial^{\,}_{x}\theta.
\end{split}
\label{eq: point defect d=1 step 1}
\end{equation}
The subscript 0 refers to the semi-classical
Green function~(\ref{eq: def semi classical G}).
We use the identity
\begin{equation}
\partial^{\,}_{\omega}\mathcal{G}^{-1}= 
\mathrm{i},
\end{equation}
introduce the family of indices labeled by $\nu$,
\begin{equation}
\nu^{\,}_{1},\nu^{\,}_{2},\nu^{\,}_{3}=0,1,2,
\end{equation}
and the Euclidean 3-momentum
\begin{equation}
K^{\,}_{\nu}:=(\omega,p,\theta). 
\end{equation}
With the help of the manipulations 
made between Eqs.%
~(\ref{appeq: gradient expansion compared to induced current if theta}) 
and 
(\ref{appeq: proof d chen number}),
it is possible to re-write
Eq.~(\ref{eq: point defect d=1 step 1}) as
\begin{equation}
\begin{split}
Q^{\,}_{\mathrm{adia}}=&\,
-\frac{1}{24\pi^{2}}
\int\mathrm{d}\omega
\int\limits_{0}^{2\pi}\mathrm{d}p
\int\limits_{0}^{2\pi}\mathrm{d}\theta\
\epsilon^{\,}_{\nu^{\,}_{1}\nu^{\,}_{2}\nu^{\,}_{3}}
\\
&\,
\times
\mathrm{tr}^{\,}_{R}
\Big( 
\mathcal{G}\partial^{\,}_{\nu^{\,}_{1}}\mathcal{G}^{-1}
\mathcal{G}\partial^{\,}_{\nu^{\,}_{2}}\mathcal{G}^{-1}
\mathcal{G}\partial^{\,}_{\nu^{\,}_{3}}\mathcal{G}^{-1}
\Big)^{\,}_{0}.
\end{split}
\label{eq: Chern id d=1}
\end{equation}
Equation~(\ref{eq: Chern id d=1})
is \textit{the first Chern number}.%
~\cite{Volovik88}
Thus, the conserved quasiparticle charge 
$Q^{\,}_{\mathrm{adia}}$
obeying Eq.~(\ref{eq: Chern id d=1})
takes integer values.
We defer to Sec.%
~\ref{sec: Physical interpretation of the adiabatic approximation}
and Appendix~\ref{appsubsec: Dirac fermions in d=1}
for a more detailed discussion of the connection between
\textit{the first Chern number}
and the number of unoccupied zero modes.

\subsection{
Chiral $\mathsf{p}$-wave superconductor when $d=2$
           }
\label{sec: Chiral mathsf{p}-wave superconductor when d=2}

\subsubsection{
Definition
           }

One of the main results of this paper consists in applying the
counting formula%
~(\ref{eq: master counting formula bis})
to the two-dimensional chiral $\mathsf{p}$-wave
superconductor with the single-particle BdG Hamiltonian
\begin{equation}
\begin{split}
&
\mathcal{H}^{\mathrm{BdG}}_{p^{\,}_{x}+\mathrm{i}p^{\,}_{y}}:=
\begin{pmatrix}
\varepsilon(\hat{\p})
& 
\frac{1}{2}
\{
\hat{p}^{\,}_{1}
-
\mathrm{i}\hat{p}^{\,}_{2},
\Delta(\x) 
\} 
\\
\frac{1}{2}
\{
\hat{p}^{\,}_{1}
+
\mathrm{i}\hat{p}^{\,}_{2},
\Delta^{*}(\x)
\} 
&
-
\varepsilon(\hat{\p})
\end{pmatrix}
\\
&
\hphantom{\mathcal{H}^{\mathrm{BdG}}_{p^{\,}_{x}+\mathrm{i}p^{\,}_{y}}:}=
\mathcal{H}^{\mathrm{BdG}\,\dag}_{p^{\,}_{x}+\mathrm{i}p^{\,}_{y}},
\\
&
\varepsilon(\hat{\p}):=
\frac{\hat{\p}^{2}}{2m}
-
\mu,
\end{split}
\label{eq: def chiral 2d p-wave mathcal H}
\end{equation}
for the case when the superconducting order parameter
supports the vorticity $n^{\,}_{\mathrm{v}}=\pm1$ at the origin
of two-dimensional Euclidean space, 
i.e., when
\begin{equation}
\Delta(\x):=
\Delta^{\ }_{0}(r)\,e^{\mathrm{i}\,n^{\ }_{\mathrm{v}}\,\theta}
\label{eq: vortex px+ipy}
\end{equation}
where $r$ and $\theta$ are the polar coordinates of 
$\x\in\mathbb{R}^{2}$.

The applicability of the counting formula%
~(\ref{eq: master counting formula bis})
follows from the antiunitary conjugation symmetry
\begin{subequations}
\label{eq: def charge conjugation for p wave SC}
\begin{equation}
\mathcal{C}^{-1}_{\mathrm{ph}}\,
\mathcal{H}^{\mathrm{BdG}}_{p^{\,}_{x}+\mathrm{i}p^{\,}_{y}}\,
\mathcal{C}^{\,}_{\mathrm{ph}}=
-
\mathcal{H}^{\,}_{p^{\,}_{x}+\mathrm{i}p^{\,}_{y}},
\label{eq: charge conjugation symmetry of px+ipy a}
\end{equation}
where 
\begin{equation}
\mathcal{C}^{\,}_{\mathrm{ph}}:= 
\rho^{\,}_{1}\,
\mathsf{K}.
\label{eq: charge conjugation symmetry of px+ipy b}
\end{equation} 
\end{subequations}
The Pauli matrices 
$\rho^{\,}_{1}$,
$\rho^{\,}_{2}$,
and
$\rho^{\,}_{3}$
encode the particle and hole block structure of
the single-particle BdG Hamiltonian%
~(\ref{eq: def chiral 2d p-wave mathcal H}).
Complex conjugation is 
denoted by $\mathsf{K}$.

For a unit vorticity, 
Read and Green have shown in Ref.~\onlinecite{Read99}
the existence of a single zero mode bound
to the vortex by solving the eigenvalue problem at zero energy.
A zero mode in a single-particle BdG Hamiltonian
realizes a Majorana fermion, for it cannot be distinguished from
its particle-hole conjugate. For an arbitrary vorticity 
$n^{\,}_{\mathrm{v}}\in\mathbb{Z}$ , 
it is shown in Refs.~\onlinecite{Tewari07} and \onlinecite{Gurarie07}
by solving the eigenvalue problem at zero energy
that the number of Majorana fermions is one if $n^{\,}_{\mathrm{v}}$ is odd
and zero otherwise.

We do not expect the adiabatic approximation
to the counting formula%
~(\ref{eq: master counting formula bis})
to capture this subtle parity effect
since it is only sensitive to
the net vorticity $n^{\,}_{\mathrm{v}}$
trapped in region of space much larger than
the characteristic length scale 
$\ell\gg1/\Delta^{\,}_{0}$
over which $|\delta\boldsymbol{\varphi}|$ changes by
the amount of order $\Delta^{\,}_{0}$.
The adiabatic approximation fails
to distinguish the case of a single vortex with
vorticity $n^{\,}_{\mathrm{v}}$ and $n^{\,}_{\mathrm{v}}$ vortices
with unit vorticity that are separated by a distance 
of order $\ell$.
This parity effect is a non-perturbative effect from the point
of view of the gradient expansion that we now present.

\subsubsection{
Counting zero modes
              }
\label{subsec: Counting zero modes}

To count the unoccupied zero modes induced by a vortex
in the superconducting order parameter
with the gradient expansion of Sec.%
~\ref{sec: Gradient expansion}, 
we need to simultaneously move in energy the 
zero mode and properly regulate the singularity
at the core of the vortex. Achieving both goals 
is impossible with the $2\times2$ BdG Hamiltonian
for the chiral $\mathsf{p}$-wave single-particle
Hamiltonian. On the other hand, both conditions are met
after doubling the BdG single-particle Hamiltonian
so as to introduce an auxiliary chiral-symmetry-breaking 
perturbation in two ways (denoted by the subscripts $\mp$),
\begin{subequations}
\label{eq: doubled px+ipy Hamiltonian}
\begin{equation}
\mathcal{H}^{\mathrm{aux}}_{\mp}\big(\hat{\p},\bm\phi(\x)\big)\!:=\!
\begin{pmatrix}
\mathcal{H}^{\mathrm{BdG}}_{p^{\,}_{x}+\mathrm{i}p^{\,}_{y}} 
& 
\frac{1}{2}
\rho^{\,}_{0}
\left\{
\hat{p}^{\,}_{\mp},
\phi^{\,}_{3}
(\x)
\right\}\,
\\
\frac{1}{2}
\rho^{\,}_{0}\,
\left\{
\hat{p}^{\,}_{\pm},
\phi^{\,}_{3}
(\x)
\right\}\,
& 
-\mathcal{H}^{\mathrm{BdG}}_{p^{\,}_{x}+\mathrm{i}p^{\,}_{y}}
\end{pmatrix}
\!.
\label{eq: doubled px+ipy Hamiltonian a}
\end{equation}
The short-hand notations
$\hat{p}^{\,}_{\pm}:=\hat{p}^{\,}_{x}\pm\mathrm{i}\hat{p}^{\,}_{y}$
was introduced and $\rho^{\,}_{0}$ is the $2\times2$ unit matrix. 
The triplet
\begin{equation}
\boldsymbol{\phi}(\x)\equiv
\begin{pmatrix}
\phi^{\,}_{1}(\x)
\\
\phi^{\,}_{2}(\x)
\\
\phi^{\,}_{3}(\x)
\end{pmatrix}
:=
\begin{pmatrix}
\Delta^{\,}_{1}(\x)
\\
\Delta^{\,}_{2}(\x)
\\
\phi^{\,}_{3}(\x)
\end{pmatrix}
\label{eq: doubled px+ipy Hamiltonian b}
\end{equation}
\end{subequations}
is real-valued and is made of the static superconducting order
parameter 
$\Delta(\x)\equiv\Delta^{\,}_{1}(\x)+\mathrm{i}\Delta^{\,}_{2}(\x)$ 
and of the static auxiliary 
charge-conjugation-symmetry-breaking field 
$\phi^{\,}_{3}(\x)$.

The spectrum of the pair of auxiliary single-particle
Hamiltonians (\ref{eq: doubled px+ipy Hamiltonian a})
given a uniform order parameter
$\boldsymbol{\phi}^{\,}_{0}$ in Eq.%
~(\ref{eq: doubled px+ipy Hamiltonian b})
is
\begin{equation}
 \varepsilon^{2}_{0}(\boldsymbol{p})= 
\left(
\frac{\p^{2}}{2m}
-
\mu
\right)^{2}
+ 
\boldsymbol{p}^{2}
\boldsymbol{\phi}^{2}_{0}.
\end{equation}
Conditions (i) and (ii) from Sec.~\ref{sec: Gradient expansion} 
are thus fulfilled since the Fermi surface $\boldsymbol{p}^{2}=2m\mu$
is fully gaped.

When $\phi^{\,}_{3}=0$, the auxiliary single-particle Hamiltonian
(\ref{eq: doubled px+ipy Hamiltonian}) 
represents two independent copies of the original
chiral $\mathsf{p}$-wave BdG Hamiltonian%
~(\ref{eq: def chiral 2d p-wave mathcal H}).
When $\phi^{\,}_{3}=0$, the spectrum of
$\mathcal{H}^{\mathrm{aux}}_{\mp}$
is identical to the spectrum of
$\mathcal{H}^{\mathrm{BdG}}_{p^{\,}_{x}+\mathrm{i}p^{\,}_{y}}$
up to a two-fold degeneracy arising from the doubling.
Furthermore, $\phi^{\,}_{3}=0$ implies that, in addition
to the antiunitary charge conjugation symmetry
\begin{subequations}
\begin{equation}
\mathcal{C}^{-1}_{\mathrm{ph}}\,
\mathcal{H}^{\mathrm{aux}}_{\mp}
\big(\hat{\boldsymbol{p}},\boldsymbol{\phi}(\boldsymbol{x})\big)\,
\mathcal{C}^{\,}_{\mathrm{ph}}
=
-
\mathcal{H}^{\mathrm{aux}}_{\mp}
\big(\hat{\boldsymbol{p}},\boldsymbol{\phi}(\boldsymbol{x})\big)
\end{equation}
with 
\begin{equation}
\mathcal{C}^{\,}_{\mathrm{ph}}=
\begin{pmatrix}
\rho^{\,}_{1} 
& 
0
\\
0 
& 
\rho^{\,}_{1}
\end{pmatrix}
\mathsf{K},
\end{equation}
\end{subequations}
where $\mathsf{K}$ 
denotes as before the operation of complex conjugation,
that originates from 
Eq.~(\ref{eq: def charge conjugation for p wave SC}),
there exists an auxiliary unitary charge-conjugation symmetry
\begin{subequations}
\begin{equation}
\mathcal{C}^{-1}_{\mathrm{ch}}
\mathcal{H}^{\mathrm{aux}}_{\mp}
\big(\hat{\boldsymbol{p}},\boldsymbol{\phi}(\boldsymbol{x})\big)\,
\mathcal{C}^{\,}_{\mathrm{ch}}
=
-
\mathcal{H}^{\mathrm{aux}}_{\mp}
\big(\hat{\boldsymbol{p}},\boldsymbol{\phi}(\boldsymbol{x})\big)
\end{equation}
with the generator of the auxiliary chiral transformation
\begin{equation}
\mathcal{C}^{\,}_{\mathrm{ch}}:=
\begin{pmatrix}
0 
&
\rho^{\,}_{0}
\\
\rho^{\,}_{0}
& 
0
\end{pmatrix}.
\end{equation}
\end{subequations}
Although neither
$\mathcal{C}^{\,}_{\mathrm{ph}}$
nor
$\mathcal{C}^{\,}_{\mathrm{ch}}$
are symmetries as soon as 
$\phi^{\ }_{3}\neq0$,
their product
$
\mathcal{T}^{\ }_{\mathrm{aux}}\equiv
\mathcal{C}^{\,}_{\mathrm{ph}}
\mathcal{C}^{\,}_{\mathrm{ch}}
$
is a symmetry for any $\phi^{\ }_{3}$.
The antiunitary operation
$
\mathcal{T}^{\ }_{\mathrm{aux}}
$
can be thought of as implementing an auxiliary 
time-reversal symmetry. As a result all eigenstates
of 
$\mathcal{H}^{\mathrm{aux}}_{\mp}\big(\hat{\p},\bm\phi(\x)\big)$,
including zero modes,
have a two-fold Kramers degeneracy.

As we shall see in Sec.%
~\ref{sec: Dirac single-particle mathcal{H} for any d},
Dirac Hamiltonians can also share 
unitary and antiunitary charge-conjugation symmetries.
There are differences with Sec.%
~\ref{sec: Dirac single-particle mathcal{H} for any d}
however. A first difference with Sec.%
~\ref{sec: Dirac single-particle mathcal{H} for any d}
is that
\begin{subequations}
\begin{equation}
\mathcal{C}^{-1}_{\mathrm{ch}}
\mathcal{H}^{\mathrm{aux}}_{\mp}
\big(\hat{\boldsymbol{p}},\boldsymbol{\phi}(\boldsymbol{x})\big)\,
\mathcal{C}^{\,}_{\mathrm{ch}}
=
-
\mathcal{H}^{\mathrm{aux}}_{\pm}
\big(\hat{\boldsymbol{p}},C^{\,}_{\mathrm{ch}}\boldsymbol{\phi}(\boldsymbol{x})\big)
\end{equation}
with
\begin{equation}
C^{\,}_{\mathrm{ch}}
\begin{pmatrix}
\phi^{\,}_{1}(\x)
\\
\phi^{\,}_{2}(\x)
\\
\phi^{\,}_{3}(\x)
\end{pmatrix}
\equiv
\begin{pmatrix}
\phi^{\,}_{1}(\x)
\\
\phi^{\,}_{2}(\x)
\\
-
\phi^{\,}_{3}(\x)
\end{pmatrix}.
\end{equation}
\end{subequations}
A second difference with Sec.%
~\ref{sec: Dirac single-particle mathcal{H} for any d}
is the absence of relativistic invariance.
A third difference is that Eq.%
~(\ref{eq: non-relativistic term})
is non-vanishing.

We start from the expansion%
~(\ref{eq: master formula before expanding in momenta})
of the  quasiparticle charge density induced by
the order parameter, the static triplet $\boldsymbol{\phi}$.
We compute first the contribution from $n=1$
for the pair of auxiliary Hamiltonians. 
It vanishes. The adiabatic approximation to 
the  quasiparticle charge density 
for any one of the pair of auxiliary Hamiltonians is
given by Eq.~(\ref{eq: n=2 adia appro})
\begin{equation}
\begin{split}
\rho^{\mathrm{aux}}_{\mp\mathrm{adia}}(\x)=&
\pm
\int
\frac{\mathrm{d}\omega\mathrm{d}^{2}\p}{(2\pi)^3}
\frac{
8\boldsymbol{p}^{2}
\left[\varepsilon(\p)+2\mu\right]
     }
     {
\left[
\omega^{2}
+
\varepsilon^{2}(\p)
+
\boldsymbol{p}^{2}\boldsymbol{\phi}^{2}(\x)
\right]^3}
\\
&\,
\times
\epsilon^{\,}_{\mathsf{abc}}
\Big(
\left(\d^{\,}_{1}\phi^{\,}_{\mathsf{a}}\right)
\left(\d^{\,}_{2}\phi^{\,}_{\mathsf{b}}\right)
\phi^{\,}_{\mathsf{c}}\phi^{\,}_{3}
\Big)(\x)
\\
=&
\pm
\int\frac{\mathrm{d}^{2}\p}{(2\pi)^{2}}
\frac{
3\p^{2}
\left[\varepsilon(\p)+2\mu\right]
     }
     {
2
\left[
\varepsilon^{2}(\p)
+
\boldsymbol{p}^{2}
\boldsymbol{\phi}^{2}(\x)
\right]^{5/2}
     }
\\
&\,  
\times
\epsilon^{\,}_{\mathsf{abc}}
\Big(
\left(\d^{\,}_{1}\phi^{\,}_{\mathsf{a}}\right)
\left(\d^{\,}_{2}\phi^{\,}_{\mathsf{b}}\right)
\phi^{\,}_{\mathsf{c}}\phi^{\,}_{3}
\Big)
(\x)
\\
=&
\pm
\frac{
1+\mathrm{sgn}(\mu)
     }
     {
2\pi\left|\boldsymbol{\phi}(\x)\right|^{4}
     }
\epsilon^{\,}_{\mathsf{abc}}
\Big(
\left(\d^{\,}_{1}\phi^{\,}_{\mathsf{a}}\right)
\left(\d^{\,}_{2}\phi^{\,}_{\mathsf{b}}\right)
\phi^{\,}_{\mathsf{c}}\phi^{\,}_{3}
\Big)
(\x).
\end{split}
\label{eq: explicit tracing for chiral p-wave}
\end{equation}

When $\mu<0$, Eq.~(\ref{eq: explicit tracing for chiral p-wave})
gives $\rho^{\mathrm{aux}}_{\mp}(\x)=0$. 
This is consistent with the absence of 
a normalizable zero energy solution for negative values of the
chemical potential.\cite{Read99,Tewari07,Gurarie07} 

When $\mu>0$ and $\phi^{\,}_{3}$ is independent of 
space $(\x)$, the adiabatic approximation to
the  conserved quasiparticle charge 
of the auxiliary Hamiltonian is
\begin{subequations}
\begin{equation}
Q^{\mathrm{aux}}_{\mp\mathrm{adia}}(\phi^{\,}_{3})=
\pm
\frac{1}{\pi}
\int\mathrm{d}\Theta\,
\int\limits_{0}^{\Delta^{\,}_{0}}
\mathrm{d}\rho\,\rho\,
\frac{
\phi^{2}_{3}
     }
     {
\left(
\rho^{2}
+
\phi^{2}_{3}
\right)^{2}
     },
\end{equation}
where the parametrization
\begin{equation}
\begin{pmatrix}
\phi_{1}
\\
\phi_{2}
\end{pmatrix}
=
\begin{pmatrix} 
\rho(r)\cos{\Theta(\theta)}
\\
\rho(r)\sin{\Theta(\theta)}
\end{pmatrix}
\end{equation}
\end{subequations}
is assumed for the superconducting order parameter
with $r$ and $\theta$ denoting 
the polar coordinates of $\boldsymbol{x}\in\mathbb{R}^{2}$.
In the limit in which $\phi^{\,}_{3}$ tends to zero, 
we get for the induced charge of the
auxiliary Hamiltonian
\begin{equation}
\begin{split}
Q^{\mathrm{aux}}_{\mp\mathrm{adia}}=&\,
\pm \frac{1}{2\pi}\int\,\mathrm{d}\Theta
\\
=&\,
\pm\text{ winding number in }(\phi^{\,}_{1},\phi^{\,}_{2}).
\end{split}
\end{equation}
To compute the number $N^{\mathrm{aux}}$ of unoccupied zero modes with
the counting formula%
~(\ref{eq: master counting formula bis})
induced by a vortex with vorticity unity,
we choose the charge-conjugation-symmetry-breaking
perturbation such that it shifts the zero mode in energy
above the chemical potential. In the limit
$\phi^{\,}_{3}\to 0$, we find that
\begin{equation}
N^{\mathrm{aux}}=2.
\label{eq: number of zero modes in double px+ipy}
\end{equation}
Equation~(\ref{eq: number of zero modes in double px+ipy})
implies that the adiabatic approximation for
the number $N$ of unoccupied zero modes of the original 
BdG Hamiltonian~(\ref{eq: def chiral 2d p-wave mathcal H})
induced by a vortex with unit vorticity is
\begin{equation}
N=1.
\end{equation}

\subsubsection{
Chern number
              }

Next, we compactify space%
~(\ref{eq: point defects and order parameter c})
and the order-parameter space%
~(\ref{eq: point defects and order parameter a}),
\begin{equation}
\boldsymbol{x}\in S^{2},
\qquad
\boldsymbol{\phi}(\boldsymbol{\theta})\in S^{2}\subset\mathbb{R}^{3}
\end{equation}
where we have introduced the spherical coordinates
$\boldsymbol{\theta}=(\theta^{\,}_{1},\theta^{\,}_{2})$
of the two-sphere $S^{2}\subset\mathbb{R}^{3}$.
Motivated by Eq.~(\ref{eq: n=2 adia appro}),
we separate the adiabatic approximation to
the  conserved quasiparticle charge 
of the auxiliary Hamiltonian into two contributions,
\begin{subequations}
\label{eq: breaking up Q adia in prime and prime prime}
\begin{equation}
Q^{\mathrm{aux}}_{\mathrm{adia}}=
Q^{\mathrm{aux}\prime}_{\mathrm{adia}}
+
Q^{\mathrm{aux}\prime\prime}_{\mathrm{adia}}
\label{eq: breaking up Q adia in prime and prime prime a}
\end{equation}
where
\begin{equation}
\begin{split}
Q^{\mathrm{aux}\prime}_{\mathrm{adia}}:=&\,
-
\int\limits_\omega
\int\limits_{\boldsymbol{p}\in S^{2}}
\int\limits_{\boldsymbol{x}\in S^{2}}
\mathrm{tr}^{\,}_{4}
\Big(
\mathcal{G}
\partial^{\,}_{i^{\,}_{2}}
\mathcal{G}^{-1}
\mathcal{G}
\partial^{\,}_{\mathsf{a}^{\,}_{2}}
\mathcal{G}^{-1}
\\
&\,\times
\mathcal{G}
\partial^{\,}_{\mathsf{a}^{\,}_{1}}
\mathcal{G}^{-1}
\mathcal{G}
\partial^{\,}_{\mathsf{i}^{\,}_{1}}
\mathcal{G}^{-1}
\mathcal{G}
\Big)^{\,}_{0}
\partial^{\,}_{i^{\,}_{1}}\phi^{\,}_{\mathsf{a}^{\,}_{1}}
\partial^{\,}_{i^{\,}_{2}}\phi^{\,}_{\mathsf{a}^{\,}_{2}}
\end{split}
\label{eq: breaking up Q adia in prime and prime prime b}
\end{equation}
while
\begin{equation}
\begin{split}
Q^{\mathrm{aux}\prime\prime}_{\mathrm{adia}}:=&\,
\frac{1}{2}
\int\limits_\omega
\int\limits_{\boldsymbol{p}\in S^{2}}
\int\limits_{\boldsymbol{x}\in S^{2}}
\mathrm{tr}^{\,}_{4}
\Big(
\mathcal{G}
\partial^{2}_{i^{\,}_{2}\mathsf{a}^{\,}_{2}}\mathcal{G}^{-1}
\mathcal{G}
\partial^{\,}_{\mathsf{a}^{\,}_{1}}\mathcal{G}^{-1}
\partial^{\,}_{i^{\,}_{1}}\mathcal{G}
\\
&\,
+
\partial^{\,}_{i^{\,}_{2}}\mathcal{G}
\partial^{\,}_{\mathsf{a}^{\,}_{2}}\mathcal{G}^{-1}
\mathcal{G}
\partial^{2}_{i^{\,}_{1}\mathsf{a}^{\,}_{1}}\mathcal{G}^{-1}
\mathcal{G}
\Big)^{\,}_{0}
\partial^{\,}_{i^{\,}_{1}}\phi^{\,}_{\mathsf{a}^{\,}_{1}}
\partial^{\,}_{i^{\,}_{2}}\phi^{\,}_{\mathsf{a}^{\,}_{2}}
\end{split}
\label{eq: breaking up Q adia in prime and prime prime c}
\end{equation}
\end{subequations}
and
$\mathcal{G}:=(\mathrm{i}\omega-\mathcal{H}^{\mathrm{aux}}_{\mp})^{-1}$.
The subscript 0 refers to the semi-classical
Green function~(\ref{eq: def semi classical G}).
The first contribution comes about when the order parameter decouples
from the momentum. The second contribution arises when the order
parameter and the momentum couple.

By explicit computation of the trace in Eq.%
~(\ref{eq: breaking up Q adia in prime and prime prime b}),
one verifies that this trace is fully antisymmetric in all the
indices $i^{\,}_{2},\mathsf{a}^{\,}_{2},i^{\,}_{1},\mathsf{a}^{\,}_{1}$
where the family $i$ and the family $\mathsf{a}$ of indices
are distinct with
\begin{equation}
i^{\,}_{1},i^{\,}_{2}=1,2,
\qquad
\mathsf{a}^{\,}_{1},\mathsf{a}^{\,}_{2}=1,2,3.
\end{equation}
Furthermore, the trace in Eq.%
~(\ref{eq: breaking up Q adia in prime and prime prime b})
gives an integrand with even contributions
in $\omega$ so that it does not vanish upon integration over 
$\omega$.
By explicit computation of the trace in Eq.%
~(\ref{eq: breaking up Q adia in prime and prime prime c}),
one verifies that this trace yields an integrand that is
odd in $\omega$ and thus vanishes upon integration over $\omega$,
\begin{equation}
Q^{\mathrm{aux}\prime\prime}_{\mathrm{adia}}=0.
\end{equation}

With the help of the manipulations 
made between Eqs.%
~(\ref{appeq: gradient expansion compared to induced current if theta}) 
and 
(\ref{appeq: proof d chen number}),
it is possible to write
\begin{subequations}
\label{eq: final Chern p-wave}
\begin{equation}
\begin{split}
Q^{\mathrm{aux}}_{\mathrm{adia}}=&\,
Q^{\mathrm{aux}\prime}_{\mathrm{adia}}
\\
=&\,
\frac{-\mathrm{i}(2\pi)^{2}}{60}
\int\limits_{K}
\epsilon^{\,}_{\nu^{\,}_{1}\cdots\nu^{\,}_{5}}
\mathrm{tr}^{\,}_{4}
\left(
\mathcal{G}\partial^{\,}_{\nu^{\,}_{1}}\mathcal{G}^{-1}
\cdots
\mathcal{G}\partial^{\,}_{\nu^{\,}_{5}}\mathcal{G}^{-1}
\right)^{\,}_{0}
\end{split}
\label{eq: final Chern p-wave a}
\end{equation}
where we have introduced the family of indices labeled by $\nu$
\begin{equation}
\nu^{\,}_{1},\cdots,\nu^{\,}_{5}=1,\cdots,5,
\label{eq: final Chern p-wave b}
\end{equation}
the five-momentum
\begin{equation}
K^{\,}_{\nu}=
(\omega,p^{\,}_{1},p^{\,}_{2},\theta^{\,}_{1},\theta^{\,}_{2}),
\label{eq: final Chern p-wave c}
\end{equation}
and the domain of integration
\begin{equation}
\int\limits_{K}\equiv
\int\frac{\mathrm{d}\omega}{2\pi}
\int\limits_{\boldsymbol{p}\in S^{2}}
\frac{
\mathrm{d}\Omega^{\,}_{2}(\boldsymbol{p})
     }
     {
(2\pi)^{2}
     }
\int\limits_{\boldsymbol{\theta}\in S^{2}}
\frac{
\mathrm{d}\Omega^{\,}_{2}(\boldsymbol{\theta})
     }
     {
(2\pi)^{2}
     }.
\label{eq: final Chern p-wave d}
\end{equation}
\end{subequations}
The ``surface'' element of the sphere $S^{2}$ is here denoted by
$\mathrm{d}\Omega^{\,}_{2}$. 
The subscript 0 refers to the semi-classical
Green function~(\ref{eq: def semi classical G}).
Equation~(\ref{eq: final Chern p-wave})
is \textit{the second Chern number}.%
~\cite{Qi08}
It takes integer values only.
It should be compared with Eq.~(\ref{appeq: final Chern d=2 Dirac}).
We defer to Secs.%
~\ref{sec: Dirac single-particle mathcal{H} for any d},
~\ref{sec: Physical interpretation of the adiabatic approximation}, 
and%
~\ref{appsubsec: Dirac fermions in d=2}
for a more detailed discussion of the connection between
\textit{this second Chern number}
and the number of unoccupied zero modes.

We have verified that the \textit{second Chern number}%
~(\ref{eq: final Chern p-wave})
vanishes for a $\mathsf{s}$-wave superconducting order parameter
that supports an isolated vortex with unit vorticity.

\subsection{
Dirac single-particle $\mathcal{H}^{\mathrm{Dirac}}_{d}$ for any $d$
           }
\label{sec: Dirac single-particle mathcal{H} for any d}

Our purpose here is to apply the counting formula%
~(\ref{eq: master counting formula bis}) 
to a Dirac single-particle Hamiltonian defined in the 
$d$-dimensional
Euclidean space with the coordinates (\ref{eq: def space}). 
We choose the dimensionality of the Dirac matrices to be
\begin{subequations}
\label{eq: def Dirac in d}
\begin{equation}
R=2^{d}.
\label{eq: def Dirac in d a}
\end{equation}
The integer $R$ is the smallest dimensionality compatible with an
irreducible representation of the Clifford algebra generated by
the unit $R\times R$ matrix $\openone$ 
and the traceless and Hermitian matrices
\begin{equation}
\Gamma^{\,}_{\mu}=\Gamma^{\dag}_{\mu},
\quad
\mathrm{tr}^{\,}_{R}\,\Gamma^{\,}_{\mu}=0,
\quad
\left\{\Gamma^{\,}_{\mu},\Gamma^{\,}_{\nu}\right\}=
2\delta^{\,}_{\mu,\nu}\openone,
\label{eq: def Dirac in d b}
\end{equation}
where $\mu,\nu=1,\cdots,2d+1$.~\cite{Zinnjustin-textbook} 
We choose the single-particle Dirac Hamiltonian such that 
a $R$-dimensional spinor is coupled to a $(d+1)$-tuplet of real-valued static
Higgs fields 
\begin{equation}
\boldsymbol{\phi}(\boldsymbol{x})\in\mathbb{R}^{d+1},
\qquad
\boldsymbol{x}\in\mathbb{R}^{d}.
\label{eq: def Higgs field for Dirac}
\end{equation}
Accordingly,
\begin{equation}
\mathcal{H}^{\mathrm{Dirac}}_{d}
\big(\hat{\boldsymbol{p}},\boldsymbol{\phi}(\boldsymbol{x})\big):=
\sum_{i=1}^{d}
\Gamma^{\,}_{i} 
\hat{p}^{\,}_{i}
+
\sum_{\mathsf{a}=1}^{d+1}
\Gamma^{\,}_{d+\mathsf{a}}
\phi^{\,}_{\mathsf{a}}(\boldsymbol{x}).
\label{eq: def Dirac Hamiltonian in any d}
\end{equation}
\end{subequations}
The decomposition~(\ref{eq: varphi = varphi0 + delta varphi})
is generalized to
\begin{equation}
\boldsymbol{\phi}(\boldsymbol{x})=
\boldsymbol{\phi}^{\,}_{0}
+
\delta\boldsymbol{\phi}(\boldsymbol{x})
\end{equation}
and is assumed to hold together with conditions (iii) from
Sec.~\ref{sec: Gradient expansion}, once we have established the
existence of charge-conjugation symmetry.
Conditions (i) and (ii) follow from the spectrum
\begin{equation}
\varepsilon^{2}_{0}(\boldsymbol{p})=
\boldsymbol{p}^{2}
+
\boldsymbol{\phi}^{2}_{0}
\end{equation}
of 
$
\mathcal{H}^{\mathrm{Dirac}}_{d}
\big(\hat{\boldsymbol{p}},\boldsymbol{\phi}^{\,}_{0}\big)
$.

The Dirac Hamiltonian~(\ref{eq: def Dirac Hamiltonian in any d}) 
has the charge-conjugation symmetry
\begin{subequations}
\label{eq: chiral sym H Dirac}
\begin{equation}
\mathcal{H}^{\mathrm{Dirac}}_{d}=
-
\Gamma^{\,}_{d+\mathsf{a}}
\mathcal{H}^{\mathrm{Dirac}}_{d}
\Gamma^{\,}_{d+\mathsf{a}}
\label{eq: choice of chiral symmetry for Dira}
\end{equation}
for any $\mathsf{a}=1,\cdots,d+1$ as soon as 
$\phi^{\,}_{\mathsf{a}}=0$ everywhere in space.  
Without loss of generality, we choose the generator of the so-called
chiral symmetry%
~(\ref{eq: choice of chiral symmetry for Dira})
to be $\Gamma^{\,}_{2d+1}$, i.e., we identify the
charge-conjugation symmetry~(\ref{eq: def second quantized H c})
with the unitary operation represented by
\begin{equation}
\mathcal{C}^{\,}_{\mathrm{ch}}:=\Gamma^{\,}_{2d+1}=
(-\mathrm{i})^{d}\,
\Gamma^{\,}_{1}\cdots\Gamma^{\,}_{2d}.
\label{eq: def Gamma 2d+1}
\end{equation}
\end{subequations}
Equation (\ref{eq: chiral sym H Dirac})
follows from $2d+1$ being an odd integer.%
~\cite{Zinnjustin-textbook}
With this choice, the vector-valued order parameter%
~(\ref{eq: def vector valued order parameter})
is
\begin{equation}
\varphi^{\,}_{a}:=\phi^{\,}_{a},
\qquad 
a=1,\cdots,d,
\end{equation}
i.e.,
\begin{equation}
D=d.
\end{equation}

It is shown in Appendix%
~\ref{appsec: Time-reversal symmetry of  H Dirac}
that there exists a symmetric matrix 
$
\mathcal{T}^{\,}_{d}
$
in the Clifford algebra%
~(\ref{eq: def Dirac in d b}) 
that implements the time-reversal symmetry 
\begin{equation}
\mathcal{H}^{\mathrm{Dirac}}_{d}=
\mathcal{T}^{-1}_{d}\,
\mathcal{H}^{\mathrm{Dirac}\,*}_{d}\,
\mathcal{T}^{\,}_{d}.
\label{eq: TRS 2d Dirac H}
\end{equation}

When 
\begin{equation}
\phi^{\,}_{d+1}=0,
\end{equation}
the symmetries
(\ref{eq: choice of chiral symmetry for Dira})
and
(\ref{eq: TRS 2d Dirac H})
of the Dirac Hamiltonian%
~(\ref{eq: def Dirac Hamiltonian in any d})
can be combined into the antiunitary operation
\begin{equation}
\mathcal{C}^{\,}_{\mathrm{ph}}:=
\mathcal{T}^{\,}_{d}\,
\mathsf{K}\,
\mathcal{C}^{\,}_{\mathrm{ch}}
\end{equation}
that implements a particle-hole transformation
and under which the Dirac Hamiltonian%
~(\ref{eq: def Dirac Hamiltonian in any d})
is invariant,
\begin{equation}
\mathcal{H}^{\mathrm{Dirac}}_{d}=
-
\left(
\mathcal{T}^{\,}_{d}
\Gamma^{\,}_{2d+1}
\right)^{-1}\,
\mathcal{H}^{\mathrm{Dirac}\,*}_{d}\,
\left(
\mathcal{T}^{\,}_{d}
\Gamma^{\,}_{2d+1}
\right).
\end{equation}

We present our main result for Dirac fermions
in three steps. The reader is referred to
Appendix~\ref{appsec: Zero modes of H Dirac}
for their detailed derivations when $d=1$ and $d=2$,
and a sketch of the derivation when $d>2$.

First, the adiabatic approximation to the counting formula%
~(\ref{eq: master counting formula bis}) 
is determined by the 
 conserved quasiparticle adiabatic charge
$Q^{\,}_{\mathrm{adia}}$.
However, $Q^{\,}_{\mathrm{adia}}$ can be computed without
assuming charge-conjugation symmetry. It is only when
using the counting formula%
~(\ref{eq: master counting formula bis})
that $Q^{\,}_{\mathrm{adia}}$ must be restricted to a
charge-conjugation-symmetric configuration of the Higgs
order parameter~(\ref{eq: point defects and order parameter a}). 
After relaxing the condition of charge-conjugation symmetry, 
we find
\begin{subequations}
\label{eq: main result on Dirac fermions}
\begin{equation}
\begin{split}
Q^{\,}_{\mathrm{adia}}=&\,
\mathrm{Chern}^{\,}_{d}
\\
\equiv&\,
\frac{
(-1)^{d(d-1)/2}(-\mathrm{i})^{d+1}d!
     }
     {
(2d+1)!
     }
(2\pi)^{d}
\int\limits_{K}
\epsilon_{\nu^{\,}_{1}\cdots\nu^{\,}_{2d+1}}
\\
&\,\times
\mathrm{tr}^{\,}_{R}
\left(
\mathcal{G}\partial^{\,}_{\nu^{\,}_{   1}}\mathcal{G}^{-1}
\cdots
\mathcal{G}\partial^{\,}_{\nu^{\,}_{2d+1}}\mathcal{G}^{-1}
\right)^{\,}_{0}
\end{split}
\label{eq: main result on Dirac fermions a}
\end{equation}
if we compactify momentum space and the target manifold, 
i.e., we have introduced the Euclidean momentum
\begin{equation}
K^{\,}_{\nu}:=
(\omega,p^{\,}_{1},\cdots,p^{\,}_{d},
\theta^{\,}_{1},\cdots,\theta^{\,}_{d})
\label{eq: main result on Dirac fermions b}
\end{equation}
\end{subequations}
with
$(p^{\,}_{1},\cdots,p^{\,}_{d})\in S^{d}$
and the spherical coordinates
$\boldsymbol{\theta}\equiv
(\theta^{\,}_{1},\cdots,\theta^{\,}_{d})$
with
$\boldsymbol{\phi}(\boldsymbol{\theta})
\in S^{d}\subset\mathbb{R}^{d+1}$.
The last inclusion
serves to emphasize that the coordinates on the $d$-sphere
in order-parameter space
can involve a sizable breaking of the conjugation symmetry
through 
$\phi^{\,}_{d+1}(\boldsymbol{\theta})\equiv
\phi(\boldsymbol{\theta})$.
The short-hand notation 
$\int\limits_{K}$
stands for the integral over 
$\mathbb{R}\times S^{d}\times S^{d}$
defined in Eq.%
~(\ref{appeq: def integral in spherical coordinateS}).
The Euclidean single-particle Green function 
$
\mathcal{G}=
\left(
\mathrm{i}\omega-\mathcal{H}^{\mathrm{Dirac}}_{d}
\right)^{-1}
$ 
is the one for the Dirac Hamiltonian.
The subscript 0 refers to the semi-classical
Green function~(\ref{eq: def semi classical G}).
The second equality in Eq.%
~(\ref{eq: main result on Dirac fermions a})
\textit{defines the $d$-th Chern number} in term of
these Dirac Euclidean single-particle Green functions.
It takes integer values and, conversely, to each integer 
$N^{\,}_{\mathrm{hedgehog}}$ 
there corresponds a static 
$\boldsymbol{\phi}^{\,}_{N^{\,}_{\mathrm{hedgehog}}}(\boldsymbol{x})\in S^{d}$
for which $Q^{\,}_{\mathrm{adia}}=N^{\,}_{\mathrm{hedgehog}}$. We call
$\boldsymbol{\phi}^{\,}_{N^{\,}_{\mathrm{hedgehog}}}(\boldsymbol{x})\in S^{d}$
a hedgehog with \textit{the $d$-th Chern number}
$N^{\,}_{\mathrm{hedgehog}}$.

Second, if we relax the condition that 
$\boldsymbol{\phi}(\boldsymbol{\theta})\in S^{d}$
and replace it with 
Eq.~(\ref{eq: point defects and order parameter b}) 
instead, then the local quasiparticle current
induced by the adiabatic variations of the Higgs fields 
in space $(\boldsymbol{x})$ and time $(x^{\,}_{0})$
is
\begin{subequations}
\begin{equation}
j^{\nu}:=
\frac{
\Omega^{-1}_{d}
     }
     {
|\boldsymbol{\phi}|^{d+1}
     }
\frac{
\epsilon^{\nu\nu^{\,}_{1}\cdots\nu^{\,}_{d}}
     }
     {
d!
     }
\epsilon^{\,}_{\mathsf{a}\mathsf{a}^{\,}_{1}\cdots\mathsf{a}^{\,}_{d}}
\phi^{\,}_{\mathsf{a}}
\partial^{\,}_{\nu^{\,}_{1}}\phi^{\,}_{\mathsf{a}^{\,}_{1}}
\cdots
\partial^{\,}_{\nu^{\,}_{d}}\phi^{\,}_{\mathsf{a}^{\,}_{d}}.
\label{eq: current Dirac any d}
\end{equation}
It obeys the continuity equation
\begin{equation}
\partial^{\,}_{\nu}j^{\nu}=0.
\end{equation}
Here, 
\begin{equation}
|\boldsymbol{\phi}|:=
\sqrt{
\phi^{2}_{1}+\cdots+\phi^{2}_{d+1}
     },
\label{eq: norm Higgs field}
\end{equation}
\end{subequations}
the $(d+1)$ indices
$\nu,\nu^{\,}_{1},\cdots,\nu^{\,}_{d}$
run over space $(\boldsymbol{x})$ and time $(x^{\,}_{0})$, 
and the $(d+1)$ indices
$\mathsf{a}^{\,}_{1},\cdots,\mathsf{a}^{\,}_{d+1}$
run over the $(d+1)$ components of the vector-valued Higgs field%
~(\ref{eq: def Higgs field for Dirac}). The Minkowski metric
is used in space ($\boldsymbol{x}$) and time ($x^{\,}_{0}$).
Finally, $\Omega^{\,}_{d}$ denotes
the area of the $d$-sphere $S^{d}$.
Equation~(\ref{eq: current Dirac any d})
was obtained by Goldstone an Wilczek
in Ref.~\onlinecite{Goldstone81}
when $d=1$ and $d=3$ 
and by Jaroszewicz
in Ref.~\onlinecite{Jaroszewicz84}
when $d=2$.
The conserved quasiparticle charge
induced by a hedgehog configuration is the integer
\begin{equation}
\begin{split}
Q^{\,}_{\mathrm{adia}}=&\,
\int \mathrm{d}^{d}\boldsymbol{x}\,
j^{0}_{\mathrm{adia}}(\boldsymbol{x})
\\
=&\,
\frac{1}{\Omega^{\,}_{d}}
\int\limits_{S^{d}} 
\mathrm{d}\Omega^{\,}_{d}
\\
=&\,
N^{\,}_{\mathrm{hedgehog}}\in\mathbb{Z}.
\end{split}
\end{equation}
The ``surface'' element (``area'') of the sphere $S^{d}$ 
is here denoted by
$\mathrm{d}\Omega^{\,}_{d}$ ($\Omega^{\,}_{d}$).

Third, we need to define a point defect that is 
compatible with the charge-conjugation symmetry
and consistent with the adiabatic approximation
(see Sec.%
~\ref{sec: Physical interpretation of the adiabatic approximation} 
for a discussion of the latter caveat).
A charge-conjugation symmetric point defect is a half hedgehog, 
i.e.,  a static Higgs $(d+1)$-tuplet%
~(\ref{eq: point defects and order parameter a}) 
that satisfies  
Eq.~(\ref{eq: point defects and order parameter b}) 
and wraps the $d$-sphere $S^{d}$ 
the half integer $\pm1/2$ number of times
(a domain wall in $d=1$, a vortex with unit vorticity in $d=2$, etc).
The counting formula%
~(\ref{eq: master counting formula bis})
predicts that there is one
unoccupied zero modes for a half-hedgehog
provided we choose the sign of the infinitesimal 
conjugation-symmetry-breaking $\phi$ in Eq.%
~(\ref{eq: point defects and order parameter a}) 
to be opposite to the sign of $\pm1/2$.

\begin{figure}
\includegraphics[angle=0,scale=0.3]{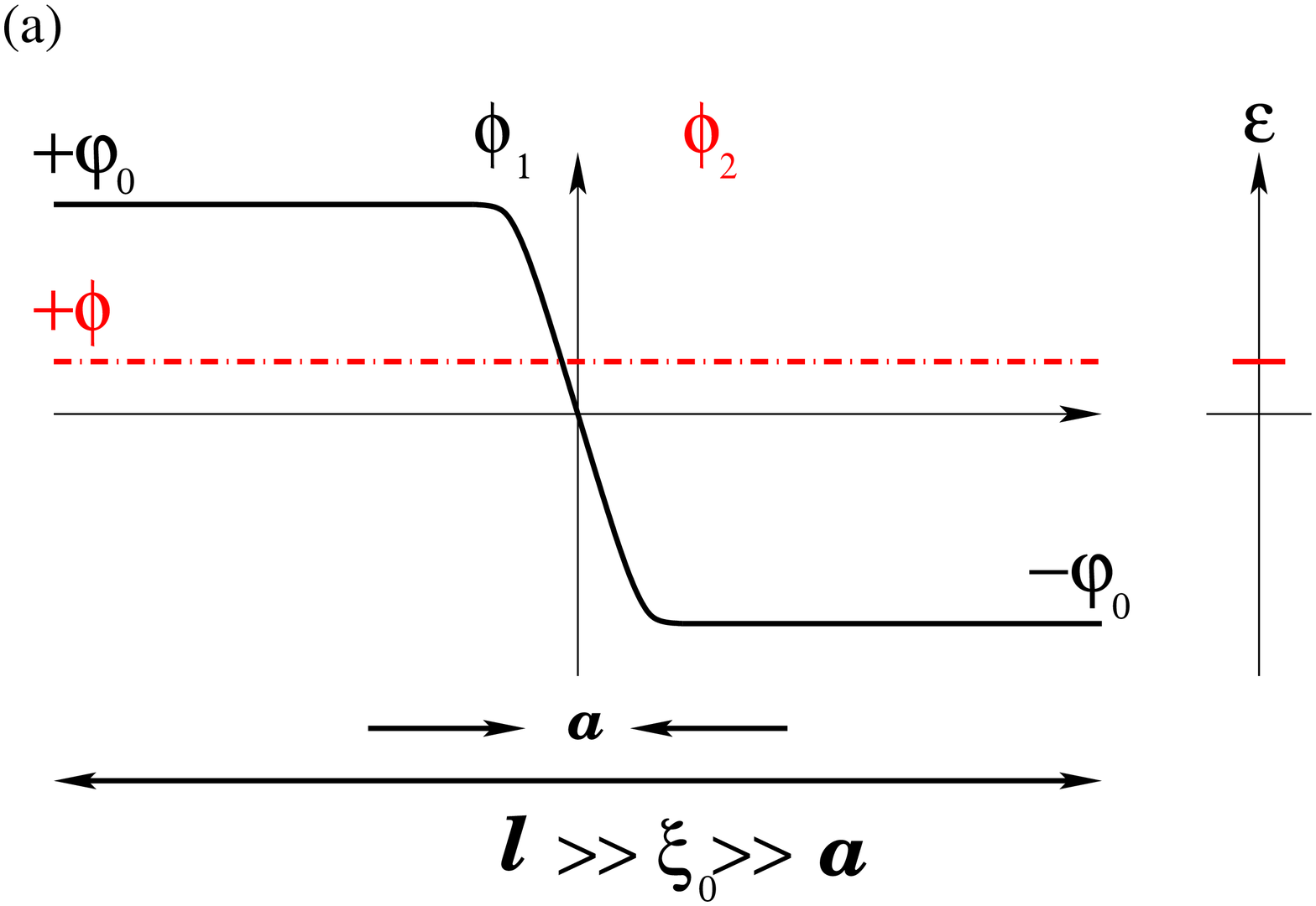}
\includegraphics[angle=0,scale=0.3]{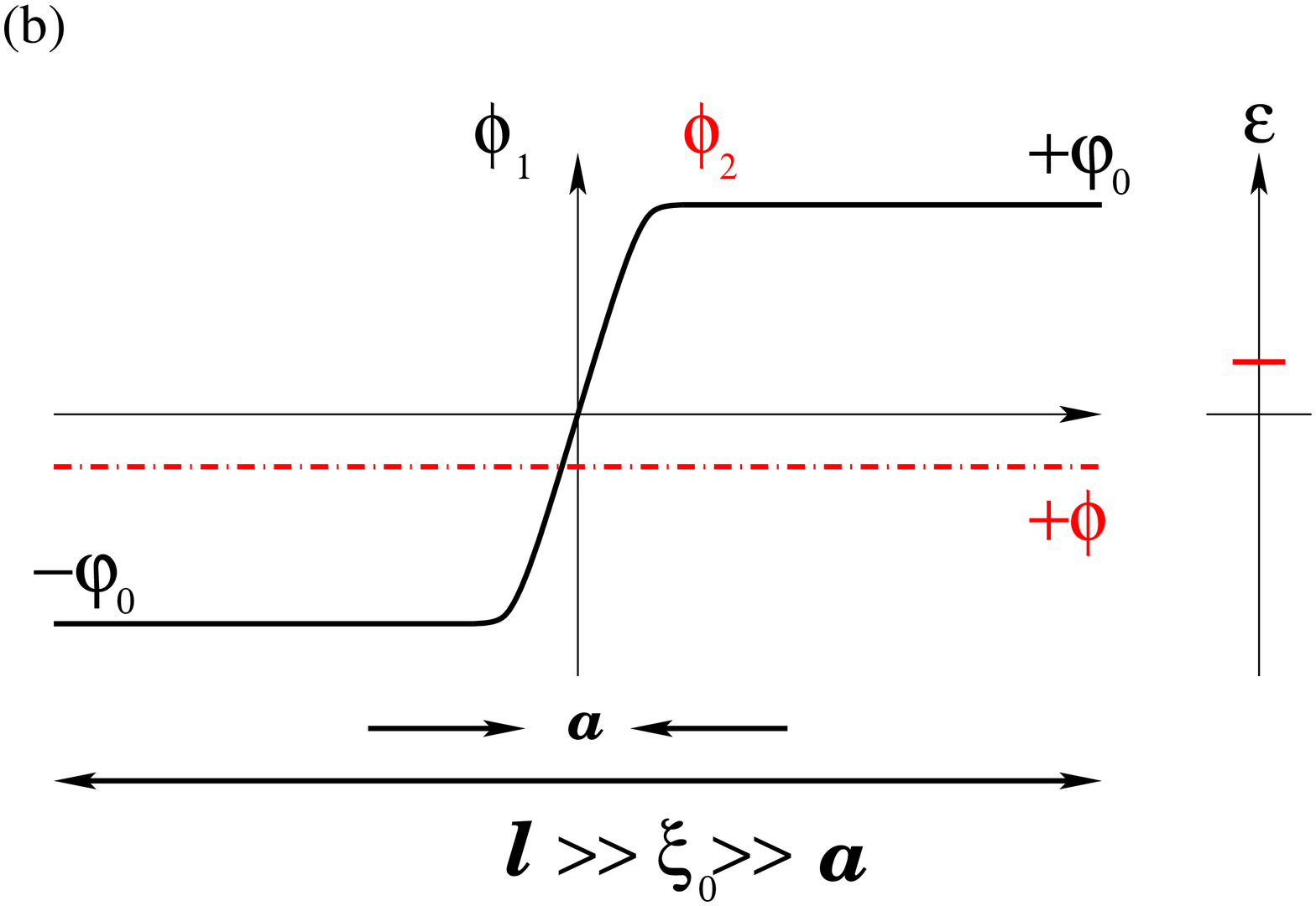}
\includegraphics[angle=0,scale=0.3]{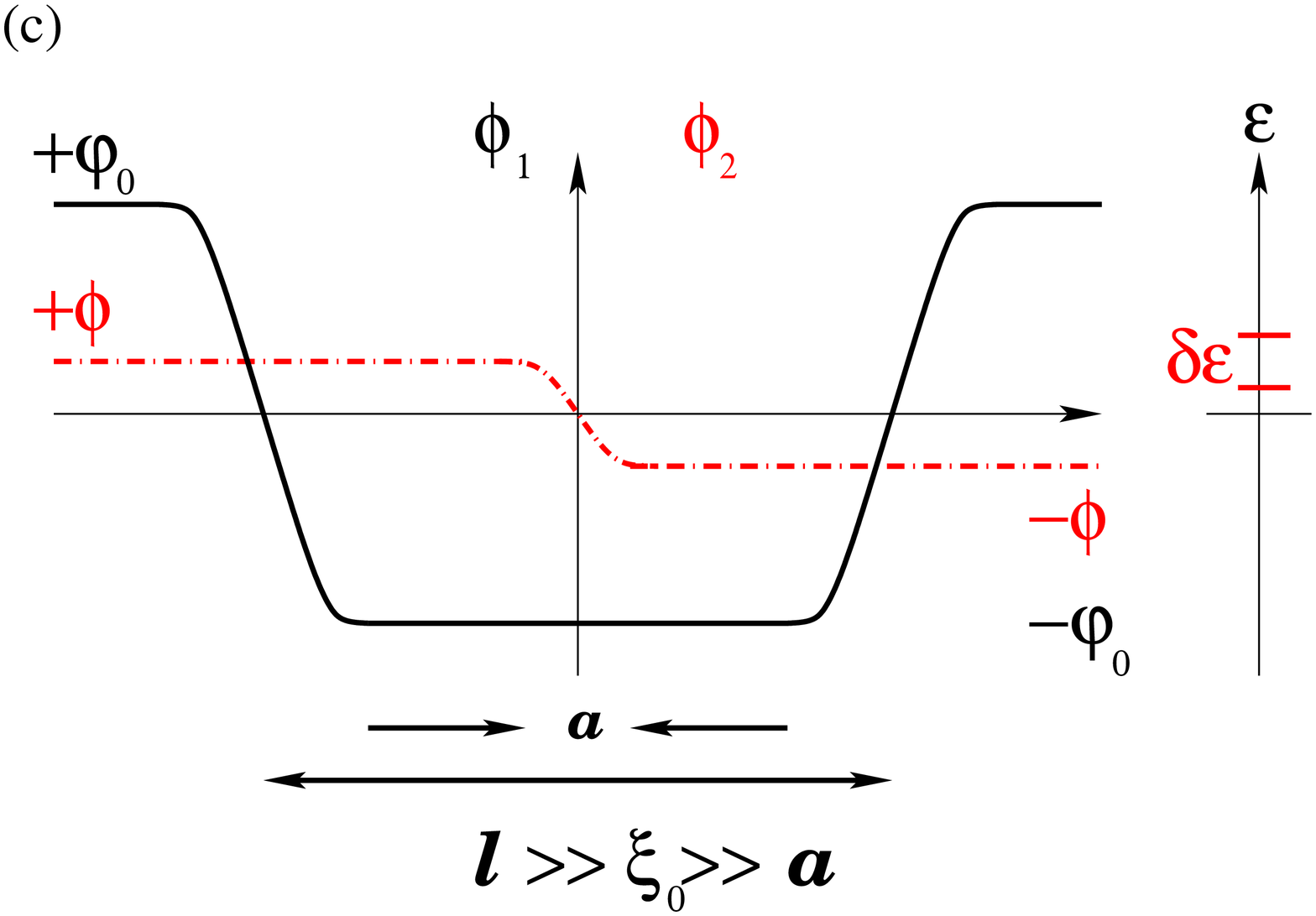}
\caption{(Color online)
The dependence on space ($x$) of the Higgs doublet in Eq.%
~(\ref{eq: polyacetilene Hamiltonian})
is depicted in panels (a), (b), and (c) together with the number and position
of the zero modes along the single-particle energy eigenvalue axis 
$\varepsilon$. The magnitude of the asymptotic value of the order parameter 
$\phi^{\,}_{1}$ is $\varphi^{\,}_{0}>0$. The magnitude of the asymptotic value of
the charge-conjugation symmetry $\phi^{\,}_{2}$ is $\phi>0$. The size of the
point defect, a domain wall, is $\mathfrak{a}$. 
The exponential decay of the envelope of the
bound state is controlled by the length scale 
$\xi^{\,}_{0}=1/|\boldsymbol{\phi}^{\,}_{0}|$.
The characteristic length $\ell$ is defined in Eq.%
~(\ref{eq: def ell from discussion sec}). 
        }
\label{Fig: domain walls}
\end{figure}

The \textit{$d$-th Chern number}%
~(\ref{eq: main result on Dirac fermions})
is non-vanishing for any single-particle
Hamiltonian topologically equivalent to
the Dirac Hamiltonian%
~(\ref{eq: def Dirac in d}).

\subsection{
Physical interpretation of the adiabatic approximation
           }
\label{sec: Physical interpretation of the adiabatic approximation} 

The relevant length scales in $d$-dimensional space are:
(1) 
the characteristic linear size $\mathfrak{a}$ of a point defect,
(2)
the characteristic linear size $\xi^{\,}_{0}=1/\Delta^{\,}_{0}$ 
of the support in space of the zero modes bound to the point defect,
(3)
the characteristic length $\ell$
over which the variation of the order parameter $\boldsymbol{\phi}$
is of order of the gap $2\Delta^{\,}_{0}$,
\begin{equation}
\left|\boldsymbol{\nabla}\boldsymbol{\phi}\right|\ell=\Delta^{\,}_{0},
\label{eq: def ell from discussion sec}
\end{equation}
(4)
and the linear size $R$ of the region of space in which the
conserved quasiparticle charge $Q$ 
and the total number $N$ of unoccupied zero modes is
to be computed (measured).
The gradient approximation relies on the hierarchy
\begin{equation}
\mathfrak{a}<
\xi^{\,}_{0}\ll
\ell\ll
R.
\end{equation}
In the gradient approximation, all microscopic data for length
scales smaller than $\ell$ are dispensed with. This fact dictates
how to properly interpret the results from the gradient 
approximation when point defects can be assigned an additive 
label $q^{\,}_{\mathrm{def}}$. Here,
the subscript stands for point defect.

When $d=1$, $q^{\,}_{\mathrm{def}}\equiv q^{\,}_{\mathrm{dw}}$,
with $q^{\,}_{\mathrm{dw}}=+1$ representing a domain wall and
$q^{\,}_{\mathrm{def}}=-1$ representing an antidomain wall.
Observe that we can exchange the terminology domain and antidomain
wall, for there is no absolute notion of a positive or negative
$q^{\,}_{\mathrm{dw}}$.
When $d=2$, $q^{\,}_{\mathrm{def}}\equiv q^{\,}_{\mathrm{vor}}\in\mathbb{Z}$
encodes the vorticity of the vortex. In either case,
these charges can be thought of as ``classical Coulomb charges''.

Any two point defects with the labels 
$q^{(1)}_{\mathrm{def}}$  and $q^{(2)}_{\mathrm{def}}$ 
within a distance of order $\mathfrak{a}$
of each other fuse into a point defect with the label
\begin{equation}
q^{(1+2)}_{\mathrm{def}}=
q^{(1)}_{\mathrm{def}} 
+
q^{(2)}_{\mathrm{def}}.
\end{equation}
When $d=1$, domain walls are necessarily ordered.
Consecutive domain walls necessarily carry the opposite 
label. Fusion of two domain walls in $d=1$ 
necessarily results in their annihilation.
When $d=2$, vortices can fuse to yield larger vorticities.

When $d=1$, we have shown in Sec.%
~\ref{appsubsec: Dirac fermions in d=1}
that the number of unoccupied zero modes $N$ 
bound to a single domain wall in a large region of size $R$
is $N=1$ in the adiabatic approximation.
This result agrees with
solving the differential equation for the zero mode
once all microscopic data have been supplied.%
~\cite{Jackiw76}
However, we have also shown in Sec.%
~\ref{subsec: Generic single-particle mathcal{H} when d=1}
and Sec.%
~\ref{appsubsec: Dirac fermions in d=1}
that the number of unoccupied zero modes $N$ 
in a large region of size $R$ can be made to be 
an arbitrary integer number in the adiabatic approximation,
in apparent contradiction with the fact that the net number of domain
walls is either 0 or 1 along any finite segment of the line.
This is explained with the help of Fig.~\ref{Fig: domain walls} 
as follows.

Occupying or leaving empty any single-particle level is
a physical operation forcing a fermionic quasiparticle
in or out of this single-particle level. This physical process
is, for zero modes, achieved by the local sign of
the charge-conjugation-symmetry-breaking $\phi^{\,}_{2}$.
In Fig.~\ref{Fig: domain walls} (a), 
we plot as a function of position $x$ along
the line the dependence of the order parameter
$\phi^{\,}_{1}$ and of the charge-conjugation-symmetry-breaking
$\phi^{\,}_{2}$. Given the shape of the domain wall
and its asymptotic value $\varphi^{\,}_{0}$, we choose
the sign $\phi^{\,}_{2}(x)=+\phi$ so that the zero mode
bound to the domain wall is shifted to a positive energy.
If the chemical potential is chosen to be at 0, then
the zero mode is unoccupied. The zero mode bound to an 
antidomain wall remains unoccupied if we choose 
the sign $\phi^{\,}_{2}(x)=-\phi$
as depicted in Fig.~\ref{Fig: domain walls}(b).
In Fig.~\ref{Fig: domain walls}(c) 
a domain wall is followed by an antidomain wall
a distance $\ell$ apart. The closing of the gap 
$|\boldsymbol{\phi}|$ at the center
of each domain walls when $\phi^{\,}_{1}=0$ 
is avoided by the charge-conjugation-symmetry-breaking field
saturating to its asymptotic magnitude $\phi$. A phase twist
in $\phi$ by $\pi$ half way between the two consecutive domain
walls insures that the zero-modes are shifted to positive energies
and thus remain unoccupied. Gap closing at this phase twist
is again avoided because $\phi^{\,}_{1}$ has healed to its 
asymptotic value. Everywhere along the line, 
$|\boldsymbol{\nabla}\boldsymbol{\phi}|$ 
can thus be made as small as needed.
The energy splitting
$\delta\varepsilon$, which is of order 
$\Delta^{\,}_{0}\exp\left(-2\ell/\xi^{\,}_{0}\right)$,
is exponentially small in $\ell/\xi^{\,}_{0}$. 
In the gradient approximation,
the order of limit is $\ell/\xi^{\,}_{0}\to\infty$ first followed
by $\phi\to0$. Evidently, this order of limit does not commute
with  $\phi\to0$ first followed by $\ell/\xi^{\,}_{0}\to\infty$.
In the latter order of limits,
the unoccupied number of zero modes always vanishes.

We close the discussion of the results in one-dimensional space
from Sec.%
~\ref{subsec: Generic single-particle mathcal{H} when d=1}
by observing that since the number of unoccupied zero mode 
is an integer, all higher contributions to the gradient expansion
beyond the adiabatic (leading) order must vanish identically. 
We have verified this expectation 
explicitly for the first sub-leading order.

When $d=2$, the adiabatic approximation of Sec.%
~\ref{sec: Chiral mathsf{p}-wave superconductor when d=2}
applied to a chiral $\mathsf{p}$-superconductor
predicts that the total number of unoccupied zero modes 
in a large region of size $R$
equals in magnitude the net vorticity in this region.
On the other hand, it is known from solving the differential equation
for a single vortex of vorticity $q^{\,}_{\mathrm{vort}}$
in a $\mathsf{p}$-wave superconductor that the number of zero modes
is one if $q^{\,}_{\mathrm{vort}}$ is odd or zero otherwise.%
~\cite{Read99,Tewari07,Gurarie07} 
This is not a paradox if the adiabatic approximation
in this paper is limited to point defects each of which
bind at most a single zero mode.
We now argue that this interpretation of the adiabatic approximation
is a necessary one.

The adiabatic approximation is not sensitive to the microscopic data
on length scales smaller than $\ell$. However, it is precisely 
those microscopic data that determines if more than one zero modes
can be bound to a point defect. Consider the case of the relativistic
Dirac Hamiltonian in two-dimensional space from Sec.%
~\ref{appsubsec: Dirac fermions in d=2}.
It respects two charge-conjugation symmetries,
the chiral symmetry and the particle-hole symmetry.
Correspondingly, the order parameter $\boldsymbol{\varphi}$
can either be interpreted as a bond-order (Kekul\'e order),%
~\cite{Hou07}
in which case the electron charge is a good quantum number, 
or as an $\mathsf{s}$-wave superconductor,%
~\cite{Jackiw81b}
in which case the electron charge is not anymore a good
quantum number (the thermal quasiparticle charge is). 
In either interpretations, an index theorem guarantees that
the number of zero modes equals in magnitude the 
vorticity of the order parameter.%
~\cite{Weinberg81,footnote-index-thm} 
For example, a vorticity of two implies
that there are two zero modes. Of course, they must be orthogonal. 
Perturb now the Dirac Hamiltonian with a perturbation, 
whose characteristic energy scale is $\eta$,
that breaks the chiral symmetry but preserves the particle-hole symmetry.%
~\cite{Ryu09} 
We assume that $\eta$ is not sufficiently large to close the gap.
The delicate balance that allowed the two zero modes to be orthogonal
is destroyed by the perturbation $\eta$. 
The two zero modes split pairwise, 
one migrating to positive energy, 
the other migrating to negative energy.
This level repulsion is encoded by the microscopic data in
a region centered about the point defect of linear size $\xi^{\,}_{0}$,
a window of length scales inaccessible to the adiabatic approximation.
If we split the single vortex with vorticity two into two vortices with
vorticity one separated by a distance $\ell$, 
we can use the adiabatic approximation.
The energy shift 
induced by the chiral-symmetry-breaking
$\phi$ is of order $\phi\exp(-2\ell/\xi^{\,}_{0})$.
The level splitting induced by the particle-hole-symmetric
$\eta$ is of order $\eta\exp(-2\ell/\xi^{\,}_{0})$.
They vanish in the adiabatic limit
$\ell/\xi^{\,}_{0}\to\infty$ first, $\phi\to0$ second.

The adiabatic approximation saturates the number of unoccupied zero modes
$N$. Consistency demands that higher-order corrections vanish identically.
This suggests that the adiabatic approximation cannot capture the
parity effect by which a charge-conjugation-symmetric
perturbation splits pairwise the degeneracy of zero modes. 
Such a parity effect is an essential singularity 
for the adiabatic expansion presented in this paper.

\section{
Conclusion
        }
\label{sec: Conclusion}

In this paper, we have derived a procedure to count the
zero-energy eigenvalues of a single-particle Hamiltonian 
$\mathcal{H}\big(\hat{\boldsymbol{p}},
\boldsymbol{\varphi}(\boldsymbol{x})\big)$ 
that possesses a charge-conjugation symmetry,
when the order parameter $\boldsymbol{\varphi}(\boldsymbol{x})$
supports point defects. This procedure is based on counting the
conserved quasiparticle charge $Q$ induced by a point defect
of $\boldsymbol{\varphi}(\boldsymbol{x})$.

The conserved quasiparticle charge $Q$ exists because of the
Hermiticity of $\mathcal{H}\big(\hat{\boldsymbol{p}},
\boldsymbol{\varphi}(\boldsymbol{x})\big)$. 
For instance, in problems for which the electric charge is a good quantum
number, the number of zero modes can be determined
from the induced electric charge near point defects. In
superconductors for which the mean-field approximation is accurate,
the conserved charge is associated to the thermal
currents carried by the Bogoliubov quasiparticles.

Irrespective of the origin of the conserved quasiparticle charge 
(electric or thermal), we showed that one can use counting arguments 
similar to those appearing implicitly in Ref.~\onlinecite{Su81} 
and explicitly in
Ref.~\onlinecite{Jackiw81a} 
to compute the conserved fractional quasiparticle charge $Q$
induced by a point defect. Once one
observes that the number of unoccupied zero modes $N$ can be related
to $Q$ through $N=-2Q$, one can concentrate the
efforts into computing $Q$ near a point defect of the position-%
dependent order parameter $\boldsymbol{\varphi}(\boldsymbol{x})$. 
We carry out this procedure within a gradient expansion for smoothly
spatially varying fluctuations of the charge-conjugation-symmetric
order parameter
$\boldsymbol{\varphi}(\boldsymbol{x})$.%
~\cite{footnote: comment on spin-wave app}

The resulting expression is then applied to generic systems in
one-dimensional space, the chiral $\mathsf{p}$-wave superconductor in
two-dimensional space, and to Dirac fermions in $d$-dimensional
space. In one-dimensional space, we find that 
the number of unoccupied zero modes is related to the
first Chern number. For the $\mathsf{p}$-wave superconductor in 
two-dimensional space, it is related to the second Chern number. 
For the Dirac Hamiltonians in $d$-dimensional space, 
the number of zero modes is determined by the $d$-th
Chern number. Therefore, we can establish in a logical and
constructive way the relation between the number of zero modes induced
by a point defect and topological invariants (the Chern numbers) in a
number of cases. 

There has been a resurgence of efforts dedicated to counting
zero modes induced by defective order parameters in the 
condensed matter community.~\cite{Sato2010,Roy2010} 
Fukui and Fujiwara in
Ref.~\onlinecite{Fukui10}
have revisited Dirac fermions in up to three spatial dimensions. 
They use the chiral anomaly to carry the counting. 
Our general counting formula reproduces their and previous results, 
and extends them to arbitrary $d$ dimensions of space. 
An elegant formulation of the Dirac problem in $d$ dimensions 
in the presence of an isotropic hedgehog has been carried out by Herbut.%
~\cite{Herbut10} 
(See also Freedman \textit{et al}.\ in Ref.~\onlinecite{Freedman10}.)
The number of zero modes should not depend on deformations away from 
the isotropic case, and this result is captured by our counting formula, 
which we can express as the $d$-th Chern number. Finally, Teo and Kane 
in Ref.~\onlinecite{Teo10}
have studied defects of arbitrary dimensions $r$
coupled to noninteracting fermionic
quasiparticles with the help of the classification scheme of 
Schnyder \textit{et al} for topological band insulators and
superconductors.%
~\cite{Schnyder08,Kitaev09,Ryu10} 
They conjecture that the topological invariant for a given
symmetry class is related to the number of zero modes attached
to $r$-dimensional defects,
but cannot make any direct connection between these two
integer numbers. Our explicit construction provides this 
relation, although, only for specific examples and for point defects. 
We do not have yet a proof for generic Hamiltonians 
that have a momentum-dependent coupling to the order parameter.

In summary, we showed that one can count zero modes 
induced by point defects using a conserved
quasi-particle charge present for any Hermitian mean-field Hamiltonian
with charge-conjugation symmetry. 
This counting supports a direct relation between topological invariants 
and the number of zero modes bound to defective order parameters.

\textbf{Acknowledgments}

This work is supported in part by the DOE Grant DE-FG02-06ER46316
(C.C.) and MIT Department of Physics Pappalardo Program (Y.N.).  
C.M.\ thanks the Condensed Matter Theory Visitor's Program at Boston
University for support and acknowledges useful discussions with T.
Neupert and T. Fukui. 
Inspiring discussions with Roman Jackiw and So-Young Pi 
are acknowledged,
in particular on whether it is possible to determine
the existence of zero modes without relying on explicit solutions.

\appendix

\section{
Counting the zero modes with the conserved quasiparticle charge
        }
\label{appsec: Schrieffer's counting argument}

In this Appendix, we review an identity that relates the
total number of zero modes of any single-particle Hamiltonian
$\mathcal{H}$ with charge-conjugation symmetry
to its Euclidean Green function
$\mathcal{G}(\omega):=(\mathrm{i}\omega-\mathcal{H})^{-1}$.
This identity appears implicitly in Ref.~\onlinecite{Su81} and 
explicitly in Ref.~\onlinecite{Jackiw81a}.%
~\cite{Wilczek02} 
We will only assume that the single-particle
Hamiltonian $\mathcal{H}$ has the property that there exists 
a transformation $\mathcal{C}$ such that
\begin{equation}
\begin{split}
\mathcal{C}^{-1}\,
\mathcal{H}\,
\mathcal{C}=
-
\mathcal{H},
\end{split}
\label{eq: charge conjugation symmetry}
\end{equation}
whereby the transformation $\mathcal{C}$
is norm preserving, i.e., it can either be an unitary or 
an antiunitary transformation. We call $\mathcal{C}$
the operation of charge conjugation.

To simplify notation, we take the spectrum 
\begin{equation}
\{0\,,\,\mathrm{sgn}(n)\,\varepsilon^{\,}_{|n|}\,|\,n=\pm1,\pm2,\cdots\}
\end{equation}
of $\mathcal{H}$ 
to be discrete up to finite degeneracies of the eigenvalues.
This is the case if $\mathcal{H}$ describes the
single-particle physics of a lattice Hamiltonian. 
The spectral decomposition of
$\mathcal{H}$ thus reads
\begin{equation}
\begin{split}
\mathcal{H}=
\sum_{n=\pm1,\pm2,\cdots}
\mathrm{sgn}(n)\varepsilon^{\,}_{|n|}
|\psi_{n}\rangle\langle \psi_{n}|,
\qquad
\varepsilon^{\,}_{|n|}>0,
\end{split}
\end{equation}
with the single-particle orthonormal basis obeying
\begin{eqnarray}
\label{eq: discrete basis}
&&
\langle \psi_{n}|\psi_{n'}\rangle=\delta^{\,}_{n,n'},
\qquad
\langle \psi_{n}|\alpha\rangle=0,
\qquad
\langle\alpha|\alpha'\rangle=
\delta^{\,}_{\alpha,\alpha'},
\nonumber \\
&&
\openone=
\sum_{\alpha=1}^{{N}}
|\alpha\rangle\langle\alpha|
+
\sum_{n=\pm1,\pm2,\cdots}
|\psi_{n}\rangle\langle \psi_{n}|
\;.
\end{eqnarray}
We have assumed the existence of ${N}$ zero modes labeled
by the index $\alpha$. The relation
\begin{equation}
|\psi_{n}\rangle= \mathcal{C}|\psi_{-n}\rangle
\end{equation}
holds for any finite energy eigenstate labeled by the index 
$n=\pm1,\pm2,\cdots$ as a result of the charge-conjugation symmetry%
~(\ref{eq: charge conjugation symmetry}).

On the lattice, we denote the value of the energy
eigenfunctions at site $i$ by
\begin{equation}
\psi^{\,}_{\mathrm{sgn}(n)\varepsilon^{\,}_{|n|,i}}:=
\langle i|\psi_{n}\rangle,
\qquad
n=\pm1,\pm2,\cdots,
\end{equation}
for the finite-energy eigenvalues and 
\begin{equation}
\psi^{\,}_{\alpha,i}:=
\langle i|\alpha\rangle,
\qquad
\alpha=1,\cdots,{N},
\end{equation}
for the ${N}$ zero modes. For any two lattice sites $i$ and $j$,
\begin{equation}
\label{eq: completeness at different sites}
\begin{split}
\delta^{\,}_{i,j}=&
\langle i|j\rangle
\\
=&
\langle i| \openone |j\rangle
\\
=&\,
\sum_{n=-1,-2,\cdots}
\psi^{* }_{-{|n|},j}
\psi^{\,}_{-{|n|},i}
\\
&+
\sum_{n=-1,-2,\cdots}
\left(\mathcal{C}\psi\right)^{* }_{-{|n|},j}
\left(\mathcal{C}\psi\right)^{\,}_{-{|n|},i}
\\
&+
\sum_{\alpha=1}^{{N}}
\psi^{* }_{\alpha,j}
\psi^{\,}_{\alpha,i},
\end{split}
\end{equation}
where we have used the completeness
relation defined in 
Eq. (\ref{eq: discrete basis}).
When $i=j$, we find the local sum rule
\begin{equation}
1=
2
\sum_{n=-1,-2,\cdots}
\psi^{* }_{-{|n|},i}
\psi^{\,}_{-{|n|},i}
+
\sum_{\alpha=1}^{{N}}
\psi^{* }_{\alpha,i}
\psi^{\,}_{\alpha,i}
\label{eq: local sum rule}
\end{equation}
owing to the fact that $\mathcal{C}$ is norm preserving.

We now assume that there exists a 
single-particle Hamiltonian
\begin{equation}
\begin{split}
\mathcal{H}^{\,}_{0}=&
\sum_{n=\pm1,\pm2,\cdots}
\mathrm{sgn}(n)\varepsilon^{0}_{|n|}
|\phi^{\,}_{n}\rangle\langle \phi^{\,}_{n}|
\\
=&
-
\mathcal{C}^{-1}\,
\mathcal{H}^{\,}_{0}\,
\mathcal{C}
\end{split}
\end{equation}
such that $\mathcal{H}^{\,}_{0}$ 
\textit{does not support zero modes}.
Evidently, the local sum rule
\begin{equation}
1=
2
\sum_{n=-1,-2,\cdots}
\phi^{ *}_{-{|n|},i}
\phi^{\,}_{-{|n|},i}
\label{eq: local sum rule (0)}
\end{equation}
also applies.

After subtracting
Eq.~(\ref{eq: local sum rule (0)})
from
Eq.~(\ref{eq: local sum rule}),
the local local sum rule
\begin{equation}
\begin{split}
&
-
2
\sum_{n=-1,-2,\cdots}
\left(
\psi^{* }_{-{|n|},i}
\psi^{\,}_{-{|n|},i}
-
\phi^{ *}_{-{|n|},i}
\phi^{\,}_{-{|n|},i}
\right)
=
\\
&
\hphantom{AAAA}
\sum_{\alpha=1}^{{N}}
\psi^{* }_{\alpha,i}
\psi^{\,}_{\alpha,i}
\end{split}
\label{eq: diff local sum rule}
\end{equation}
follows. In turn,
after summing over all lattice sites and making
use of the fact that the zero modes are normalized to one,
the global sum rule
\begin{equation}
\frac{{N}}{2}=
\sum_{n=-1,-2,\cdots}
\sum_{i}
\left(
\phi^{ *}_{-{|n|},i}
\phi^{\,}_{-{|n|},i}
-
\psi^{* }_{-{|n|},i}
\psi^{\,}_{-{|n|},i}
\right)
\label{eq: global sum rule}
\end{equation}
follows.
The global sum rule is only meaningful in the thermodynamic
limit after this subtraction procedure has been taken.

\begin{figure}
\includegraphics[angle=0,scale=0.5]{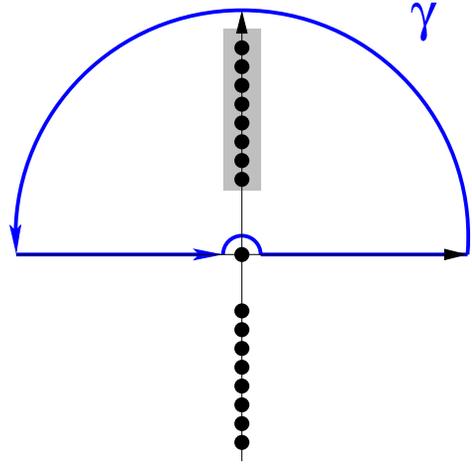}
\caption{(Color online)
Definition of the integration contour $\gamma$ that picks
up the discrete negative energy eigenvalues of the 
single-particle Hamiltonian. The shaded box represents the Fermi 
sea. The filled circles represent the discrete energy eigenvalues.
The contour $\gamma$ avoids the zero mode.
        }
\label{Fig: gamma contour}
\end{figure}

We now present the global sum rule~(\ref{eq: global sum rule})
with the help of Euclidean single-particle Green functions.
We thus define the Euclidean single-particle Green functions
\begin{subequations}
\begin{equation}
\begin{split}
\mathcal{G}(\omega):=&\,
\frac{1}{\mathrm{i}\omega-\mathcal{H}}
\end{split}
\end{equation}
and
\begin{equation}
\begin{split}
\mathcal{G}^{\,}_{0}(\omega):=&\,
\frac{1}{\mathrm{i}\omega-\mathcal{H}^{\,}_{0}}
\end{split}
\end{equation}
\end{subequations}
for any real-valued and non-vanishing $\omega$.
Next, we choose $\gamma$ to be the contour in the complex
$\omega$ plane that runs counterclockwise
along the real axis $\omega\in\mathbb{R}$, 
avoids the origin $\omega=0$ by an infinitesimal deformation into
the upper complex plane $\mathrm{Re}\,\omega>0$, and closes
through a semi-circle in the very same upper complex plane
(see Fig.~\ref{Fig: energy levels}).
With the help of the residue theorem,
it then follows that
\begin{subequations}
\label{eq:application of the residue them given gamma}
\begin{equation}
\sum_{n=-1,-2,\cdots}
\psi^{* }_{-{|n|},i}
\psi^{\,}_{-{|n|},i}
=
\int\limits_{\gamma}\frac{\mathrm{d}\omega}{2\pi}
\left\langle i\left|\mathcal{G}(\omega)\right|i\right\rangle
\end{equation}
and
\begin{equation}
\sum_{n=-1,-2,\cdots}
\phi^{* }_{-{|n|},i}
\phi^{\,}_{-{|n|},i}
=
\int\limits_{\gamma}\frac{\mathrm{d}\omega}{2\pi}
\left\langle i\left|\mathcal{G}^{\,}_{0}(\omega)\right|i\right\rangle.
\end{equation}
\end{subequations}
Equation~(\ref{eq: global sum rule}) 
can now be rewritten in the desired form
\begin{subequations}
\label{eq: global sum rule with Green}
\begin{equation}
\frac{N}{2}=
-\mathrm{Tr}\,
\int\limits_{\gamma}\frac{\mathrm{d}\omega}{2\pi}
\left[
\mathcal{G}         (\omega)
-
\mathcal{G}^{\,}_{0}(\omega)
\right].
\label{eq: global sum rule with Green a}
\end{equation}
Equation~(\ref{eq: global sum rule with Green a})
can be expressed in terms of the 
local ``lattice charge'' $\rho^{\,}_{i}$ through
\begin{equation}
\begin{split}
&
\frac{N}{2}=
-
\sum_{i} 
\rho^{\,}_{i}\equiv
-
Q,
\\
&
\rho^{\,}_{i}:=
\int\limits_{\gamma}\frac{\mathrm{d}\omega}{2\pi}
\left\langle i\left|
\left[
\mathcal{G}         (\omega)
-
\mathcal{G}^{\,}_{0}(\omega)
\right]
\right|i\right\rangle,
\end{split}
\label{eq: global sum rule with Green b}
\end{equation}
or the local ``continuum charge'' through 
``charge'' $\rho(\boldsymbol{r})$
\begin{equation}
\begin{split}
&
\frac{N}{2}=
-
\int\mathrm{d}^{d}\boldsymbol{r}\,
\rho(\boldsymbol{r})\equiv
-
Q,
\\
&
\rho(\boldsymbol{r}):=
\int\limits_{\gamma}\frac{\mathrm{d}\omega}{2\pi}
\left\langle \boldsymbol{r}\left|
\left[
\mathcal{G}         (\omega)
-
\mathcal{G}^{\,}_{0}(\omega)
\right]
\right|\boldsymbol{r}\right\rangle.
\end{split}
\label{eq: global sum rule with Green c}
\end{equation}
\end{subequations}
Hermiticity of $\mathcal{H}$
and $\mathcal{H}^{\,}_{0}$ guarantees that 
these  ``quasiparticle charge densities'' 
obey a continuity equation,
i.e., that their 
 quasiparticle charge $Q$ is conserved.

\section{
Time-reversal symmetry of 
$\mathcal{H}^{\mathrm{Dirac}}_{d}$
        }
\label{appsec: Time-reversal symmetry of  H Dirac}

We are going to prove the existence
of a time-reversal symmetry for the Dirac Hamiltonian
defined by Eq.~(\ref{eq: def Dirac in d}).
To this end, we shall introduce an auxiliary Lagrangian 
$\mathcal{L}^{\ }_{\mathrm{Dirac}}$
in $(2d+1)$-dimensional space and time.

By construction,\cite{Zinnjustin-textbook}
\begin{equation}
\begin{split}
\mathcal{L}^{\mathrm{Dirac}}_{2d+1}
(x^{\,}_{1},\cdots,x^{\,}_{2d},x^{\,}_{2d+1}):=&\,
\sum_{\nu=1}^{2d+1}
\Gamma^{\,}_{\nu}
\frac{\partial}{\partial x^{\,}_{\nu}}
\end{split}
\end{equation}
can be interpreted as the free Euclidean Dirac Lagrangian in
$(2d+1)$-dimensional Euclidean 
space $(\boldsymbol{x})$ and time $(x^{\,}_{2d+1})$. 
Observe here that $2^{d}$, where
$d=[(2d+1)/2]$ 
is the smallest integer larger or equal to the 
real number $(2d+1)/2$, is the smallest dimensionality of an irreducible
representation of a Dirac Hamiltonian in $2d$-dimensional space.%
~\cite{Zinnjustin-textbook}
In the reminder of the argument, we adopt the summation
convention over repeated
indices.

The classical action
\begin{equation}
S=
-
\int \mathrm{d}^{2d+1} x\,
\bar\psi^{\ }_{\alpha}(x)
\left(
\Gamma^{\,}_{\nu}
\frac{\partial}{\partial x^{\,}_{\nu}}
\right)^{\ }_{\alpha\beta}
\psi^{\ }_{\beta}(x)
\end{equation}
is invariant under the group $\mathrm{O}(2d+1)$ 
generated by 
\begin{equation}
\Sigma^{\ }_{\nu\nu'}:=
\frac{1}{2\mathrm{i}}
[\Gamma^{\ }_{\nu},\Gamma^{\ }_{\nu'}],
\qquad
1\leq\nu<\nu'\leq 2d+1.
\end{equation}
Consequently, there must exist
$(2d+1)$ unitary matrices
$\mathcal{P}^{\,}_{\nu}$
from the (complex) Clifford algebra such that
\begin{equation}
\mathcal{P}^{-1}_{\nu}\,
\Gamma^{\ }_{\nu'}\,
\mathcal{P}^{\,}_{\nu}=
(-)^{\delta^{\ }_{\nu,\nu'}}
\Gamma^{\ }_{\nu'},
\qquad
\nu,\nu'=1,\cdots,2d+1.
\end{equation}
Symmetry under reflection about any
direction $\nu=1,\cdots,2d+1$ in $(2d+1)$-dimensional
space and time thus becomes
\begin{equation}
\begin{split}
&
\mathcal{L}^{\mathrm{Dirac}}_{2d+1}
(\cdots,x^{\,}_{\nu-1},x^{\,}_{\nu},x^{\,}_{\nu+1},\cdots)=
\\
&\hphantom{A}
\mathcal{P}^{-1}_{\nu}\,
\mathcal{L}^{\mathrm{Dirac}}_{2d+1}
(\cdots,+x^{\,}_{\nu-1},-x^{\,}_{\nu},+x^{\,}_{\nu+1},\cdots)\,
\mathcal{P}^{\,}_{\nu}.
\end{split}
\end{equation}
There must also exist an auxiliary unitary matrix
$\mathcal{T}^{\ }_{\mathrm{aux}}$
from the (complex) Clifford algebra
under which
\begin{equation}
\begin{split}
&
\mathcal{L}^{\mathrm{Dirac}}_{2d+1}
(x^{\,}_{1},\cdots,x^{\,}_{2d},x^{\,}_{2d+1})=
\\
&\hphantom{A}
\mathcal{T}^{-1}_{\mathrm{aux}}\,
\mathcal{L}^{* }_{\mathrm{Dirac}}
(+x^{\,}_{1},\cdots,+x^{\,}_{2d},-x^{\,}_{2d+1})\,
\mathcal{T}^{\,}_{\mathrm{aux}}.
\end{split}
\end{equation}

We are now ready to define the unitary matrix
\begin{equation}
\mathcal{T}^{\,}_{d}:=
\mathcal{T}^{\,}_{\mathrm{aux}}
\mathcal{P}^{\,}_{2d+1}
\prod_{a=1}^{d}
\mathcal{P}^{\,}_{a}
\end{equation}
from the (complex) Clifford algebra.
If $\mathsf{K}$ denotes complex conjugation,
we interpret the operation
$
\mathcal{T}^{\,}_{d}\,\mathsf{K}
$
as implementing reversal of time on
the Dirac Hamiltonian%
~(\ref{eq: def Dirac Hamiltonian in any d}).
It is a symmetry, for
\begin{equation}
\mathcal{H}^{\mathrm{Dirac}}_{d}=
\mathcal{T}^{-1}_{d}\,
\mathcal{H}^{\mathrm{Dirac}\,*}_{d}\,
\mathcal{T}^{\,}_{d}.
\label{eq: TRS 2d Dirac H app}
\end{equation}

\section{
Zero modes of $\mathcal{H}^{\mathrm{Dirac}}_{d}$
        }
\label{appsec: Zero modes of H Dirac}

This Appendix provides intermediary steps for
Sec.~\ref{sec: Dirac single-particle mathcal{H} for any d}.

\subsection{
Dirac fermions in one-dimensional space
           }
\label{appsubsec: Dirac fermions in d=1}

\subsubsection{
Definition
              }
\label{appsubsubsec: Definition}

When $d=1$, Eq.~(\ref{eq: def Dirac in d})
simplifies to
\begin{equation}
\mathcal{H}^{\mathrm{Dirac}}_{1}
\big(\hat{p},\boldsymbol{\phi}(x)\big):= 
\sigma^{\,}_{3}\hat{p}
+
\sigma^{\,}_{1}\phi^{\,}_{1}(x)
+
\sigma^{\,}_{2}\phi^{\,}_{2}(x).
\label{eq: polyacetilene Hamiltonian}
\end{equation}
The $(R=2)$-dimensional representation of
the Clifford algebra is here generated
from the Pauli matrices 
$\sigma^{\,}_{1}$,
$\sigma^{\,}_{2}$,
and
$\sigma^{\,}_{3}$.
If the doublet of Higgs fields $\boldsymbol{\phi}^{\,}_{0}$
is constant through one-dimensional space ($x$),
the single-particle spectrum of
Hamiltonian (\ref{eq: polyacetilene Hamiltonian})
has a gap controlled by the Higgs components adding in quadrature,
\begin{equation}
\varepsilon^{2}_{0}(\p)
=
\p^{2} 
+ 
\bm{\phi}^{2}_{0}
=
\p^{2} 
+ 
\phi^{2}_{0\,1}
+ 
\phi^{2} _{0\,2}.
\end{equation}

The generator of the chiral symmetry of
the Dirac Hamiltonian%
~(\ref{eq: polyacetilene Hamiltonian})
is
\begin{subequations}
\begin{equation}
\mathcal{C}^{\,}_{\mathrm{ch}}=\sigma^{\,}_{2}
\end{equation}
if
\begin{equation}
\phi^{\,}_{2}=0
\end{equation}
\end{subequations}
everywhere in Euclidean space $x\in\mathbb{R}$.

The operation of time-reversal is implemented by
\begin{equation}
\mathcal{T}^{\,}_{1}\,\mathsf{K}:= 
\sigma^{\,}_{1}\,
\mathsf{K}
\end{equation}
where $\mathsf{K}$ denotes complex conjugation.
It is a symmetry of the Dirac Hamiltonian%
~(\ref{eq: polyacetilene Hamiltonian})
for any Higgs configuration 
$\phi^{\,}_{1}$ and $\phi^{\,}_{2}$.

The operation of particle-hole exchange is implemented by
\begin{equation}
\mathcal{C}^{\,}_{\mathrm{ph}}:=
\mathcal{T}^{\,}_{1}\,
\mathsf{K}\,
\mathcal{C}^{\,}_{\mathrm{ch}}=
-\mathrm{i}
\sigma^{\,}_{3}\,
\mathsf{K}
\end{equation}
and it is only a symmetry of
the Dirac Hamiltonian~(\ref{eq: polyacetilene Hamiltonian})
provided $\phi^{\,}_{2}=0$.

The discovery that this model supports zero modes was
made by Jackiw and Rebbi in Ref.~\onlinecite{Jackiw76}.
Its relevance to the physics of polyacetylene
was made by Su, Schrieffer, and Heeger in
Refs.~\onlinecite{Su79} and \onlinecite{Su80}.
In polyacetylene, the kinetic energy results from linearizing
the dispersion around the two Fermi points of a single-band 
nearest-neighbor tight-binding model at half-filling 
in the left- and right-mover basis. The Higgs field 
$\phi^{\,}_{1}\equiv\varphi^{\,}_{1}$ 
realizes a modulation of the
nearest-neighbor hopping amplitude that is mediated by phonons.
A Peierls transition opens up a single-particle electronic gap
through the breaking of the translation symmetry by one lattice spacing
down to a residual translation symmetry by two lattice spacings.
A domain wall in $\varphi^{\,}_{1}$ that interpolates between
the two possible dimer ground states binds one zero mode per 
electronic spin (which we have ignored so far). The Higgs field
$\phi^{\,}_{2}$ breaks the sublattice symmetry of the 
tight-binding model. The 
quasiparticle charge density induced by a domain wall 
with a non-vanishing charge-conjugation symmetry breaking
$\phi^{\,}_{2}$ was computed by 
Goldstone and Wilczek in Ref.~\onlinecite{Goldstone81}
and shown to vary continuously with $\phi^{\,}_{2}$.
We are going to reproduce all these results using
the adiabatic approximation~(\ref{eq: n=1 adia appro}).

\subsubsection{
Counting zero modes
              }
\label{appsubsubsec: Counting zero modes}

We start from the expansion%
~(\ref{eq: master formula before expanding in momenta a})
of the  quasiparticle charge density
which, for a Dirac Hamiltonian,
is exact and consider the contribution from $n=1$.
The adiabatic approximation to 
the  quasiparticle charge density is
\begin{equation}
\begin{split}
 \rho^{\,}_{\mathrm{adia}}(x)=&\, 
\int\frac{\mathrm{d}\omega\mathrm{d}p}{(2\pi)^{2}}
 \frac{
2\epsilon^{\,}_{\mathsf{ab}}
\left(
\partial^{\,}_{x}\phi^{\,}_{\mathsf{a}}
\right)(x)\,\phi^{\,}_{\mathsf{b}}(x)
      }
      {
\big[
\omega^{2}
+
p^{2}
+
\phi^{2}_{1}(x)
+
\phi^{2}_{2}(x)
\big]^{2}
      }
\\
=&\,
\frac{\epsilon^{\,}_{\mathsf{ab}}}{2\pi}
\frac{
\left(\partial^{\,}_{x}\phi^{\,}_{\mathsf{a}}\right)(x)\,\phi^{\,}_{\mathsf{b}}(x)
     }
     {
\left[
\phi^{2}_{1}(x)
+
\phi^{2}_{2}(x)
\right]
     }.
\end{split}
\end{equation}    
For a constant $\phi^{\,}_{2}>0$ that breaks the conjugation symmetry, 
integration over one-dimensional Euclidean space ($x$) 
gives the adiabatic approximation to the 
 conserved quasiparticle charge
\begin{equation}
\begin{split}
Q^{\,}_{\mathrm{adia}}(\phi^{\,}_{2})=&\,
\int\limits^{\phi^{\,}_{1}(x=+\infty)}_{\phi^{\,}_{1}(x=-\infty)}
\frac{\mathrm{d}\phi^{\,}_{1}}{2\pi}
\frac{
\phi^{\,}_{2}
     }
     {
\left(\phi^{2}_{1}+\phi^{2}_{2}\right)
     }
\\
=&\,
(2\pi)^{-1}
\left.
\left(
\arctan\frac{\phi^{\,}_{1}(x)}{\phi^{\,}_{2}}
\right)
\right|^{x=+\infty}_{x=-\infty}.
\end{split}
\label{eq: charge of JR before limit}
\end{equation}
The single domain wall with the asymptotic values
\begin{equation}
\phi^{\,}_{1}(\pm\infty)= 
\mp
\varphi^{\,}_{0},
\qquad
\varphi^{\,}_{0}>0,
\end{equation}
here chosen in such a way that any zero mode is shifted 
in energy by $\phi^{\,}_{2}>0$ 
\textit{above} the chemical potential, 
induces the \textit{negative} 
 conserved quasiparticle charge 
\begin{equation}
\lim_{\phi^{\,}_{2}\to0^{+}}Q^{\,}_{\mathrm{adia}}(\phi^{\,}_{2})=
-
\frac{1}{2}.
\label{eq: d=1 Q with chiral symmetry}
\end{equation}
Having restored the charge-conjugation symmetry by removing
$\phi^{\,}_{2}$ in Eq.~(\ref{eq: d=1 Q with chiral symmetry}), 
the counting formula%
~(\ref{eq: master counting formula bis})
can, in turn, be used  
to deliver the \textit{positive} number of unoccupied zero modes
\begin{equation}
N=1
\end{equation}
bound to this single domain wall.

\subsubsection{
Chern number
              }
\label{appsubsubsec: Chern number for Dirac if d=1}

Whereas the counting formula%
~(\ref{eq: master counting formula bis})
relies on the charge-conjugation symmetry,
the adiabatic approximation to the  quasiparticle
charge density does not. We are going to take advantage of this
freedom to relate the  conserved quasiparticle charge to
\textit{the first Chern number}.

To this end, 
we compactify the base space
$x\in\mathbb{R}$ to the circle $x\in S^{1}$
by imposing periodic boundary conditions.
We then
parametrize the doublet of Higgs fields
according to the polar decomposition
\begin{equation}
\boldsymbol{\phi}(x):=
\begin{pmatrix}
\phi^{\,}_{1}(x)
\\
\phi^{\,}_{2}(x)
\end{pmatrix} 
= 
m
\begin{pmatrix}
\sin\theta(x)
\\
\cos\theta(x)
\end{pmatrix}.
\end{equation}
We have thus compactified the target space
$\boldsymbol{\phi}(x)\in\mathbb{R}^{2}$
to the unit circle
$\theta(x)\in S^{1}$.

In the adiabatic approximation%
~(\ref{eq: n=1 adia appro}), 
the  conserved quasiparticle charge is
\begin{equation}
\begin{split}
Q^{\,}_{\mathrm{adia}}=&\,
\frac{1}{2(2\pi)^{2}}
\int
\mathrm{d}\omega
\int\limits_{0}^{2\pi}
\mathrm{d}p
\int\limits_{0}^{2\pi}
\mathrm{d}x\,
(-\mathrm{i})\frac{\d\theta}{\d x}
\\
&\,
\times
\mathrm{tr}
\Bigg(
\mathcal{G}
\frac{\d \mathcal{G}^{-1}}{\d p}
\mathcal{G}
\frac{\d \mathcal{G}^{-1}}{\d\theta}
\mathcal{G}
-
\mathcal{G}
\frac{\d \mathcal{G}^{-1}}{\d\theta}
\mathcal{G}
\frac{\d \mathcal{G}^{-1}}{\d p}
\mathcal{G}
\Bigg)^{\,}_{0}.
\end{split}
\label{eq: charge 1D before Chern number}
\end{equation}
The integrand can be written in a more compact and symmetric
form by observing that
\begin{equation}
\left(\frac{\partial\mathcal{G}^{-1}}{\partial\omega}\right)^{\,}_{0} =
\mathrm{i},
\qquad
\mathrm{i}\,\mathcal{G}^{\,}_{0}=
\left(
\mathcal{G}
\frac{\partial\mathcal{G}^{-1}}{\partial\omega}
\right)^{\,}_{0},
\end{equation}
and by introducing the three-momentum
\begin{equation}
K^{\,}_{\nu}\equiv
(\omega,p,\theta)\,
\in\mathbb{R}\times S^1\times S^1.
\end{equation} 
Equation (\ref{eq: charge 1D before Chern number}) 
is then nothing but \textit{the first Chern number},%
~\cite{Volovik88}
for
\begin{equation}
Q^{\,}_{\mathrm{adia}}\!=
\frac{-1}{24\pi^{2}}
\!\!\!\!\!\!\!\!
\int\limits_{\mathbb{R}\times S^1\times S^1}
\!\!\!\!\!\!\!\!
\mathrm{d}^{3}K
\epsilon^{\,}_{\mu\nu\lambda}
\mathrm{tr}
\left(
\mathcal{G}
\frac{\d \mathcal{G}^{-1}}{\d K^{\,}_{\mu}}
\mathcal{G}
\frac{\d \mathcal{G}^{-1}}{\d K^{\,}_{\nu}}
\mathcal{G}
\frac{\d \mathcal{G}^{-1}}{\d K^{\,}_{\lambda}}
\right)^{\,}_{0}. 
\label{eq: 1D charge as a Chern number}
\end{equation}
We infer that the  conserved quasiparticle charge
induced by the adiabatic winding of the Higgs doublet
around the circle takes integer values. 

Moreover, the domain wall from Sec.%
~\ref{appsubsec: Dirac fermions in d=1} 
is a half-winding of the unit circle $S^{1}$.
More precisely, evaluation of the trace in the integrand
of \textit{the first Chern number}%
~(\ref{eq: 1D charge as a Chern number})
gives the adiabatic approximation to the 
 conserved quasiparticle charge 
\begin{equation}
\begin{split}
Q^{\,}_{\mathrm{adia}}=&\, 
\frac{1}{2\pi}
\int
\mathrm{d}x\,
\frac{
\epsilon^{\,}_{\mathsf{ab}}\phi^{\,}_{\mathsf{a}}\partial^{\,}_{x}\phi^{\,}_{\mathsf{b}}
     }
     {
|\boldsymbol{\phi}|^{2}
     }
\\
=&\,
\text{winding number in } \bm{\phi}.
\end{split}
\end{equation}
The integrand is nothing but the space ($x$) and time ($t$) 
dependent quasiparticle charge density
\begin{subequations}
\begin{equation}
j^{0}:=
\frac{1}{2\pi|\boldsymbol{\phi}|^{2}}
\epsilon^{\,}_{\mathsf{ab}}\phi^{\,}_{\mathsf{a}}\partial^{\,}_{x}\phi^{\,}_{\mathsf{b}}
\end{equation}
that obeys the continuity equation~\cite{Goldstone81}
\begin{equation}
\partial^{\,}_{\nu}j^{\nu}=0
\end{equation}
with
\begin{equation}
j^{\nu}=
\frac{1}{2\pi\boldsymbol{\phi}^{2}}
\epsilon^{\nu\nu'}
\epsilon^{\,}_{\mathsf{aa}'}
\phi^{\,}_{\mathsf{a}}
\partial^{\,}_{\nu'}
\phi^{\,}_{\mathsf{a}'}.
\end{equation}
\end{subequations}

\subsection{
Dirac fermions in two-dimensional space
            }
\label{appsubsec: Dirac fermions in d=2}

\subsubsection{
Definition
              }

When $d=2$, Eq.~(\ref{eq: def Dirac in d})
simplifies to
\begin{subequations}
\label{eq: def 2D Dirac Hamiltonian}
\begin{equation}
\mathcal{H}^{\mathrm{Dirac}}_{2}
\big(\hat{p},\boldsymbol{\phi}(x)\big):= 
\alpha^{\,}_{1}
\hat{p}^{\,}_{1}
+
\alpha^{\,}_{2}
\hat{p}^{\,}_{2}
+
\beta^{\,}_{1}
\phi^{\,}_{1}
+
\beta^{\,}_{2}
\phi^{\,}_{2}
+
\beta^{\,}_{3}
\phi^{\,}_{3}.
\label{eq: graphene Hamiltonian}
\end{equation}
The $(R=4)$-dimensional representation of the
Clifford algebra can be chosen to be generated 
from the five traceless and Hermitian matrices
\begin{equation}
\Gamma^{\,}_{\nu}=
(\alpha^{\,}_{1},\alpha^{\,}_{2},
 \beta^{\,}_{1},\beta^{\,}_{2},\beta^{\,}_{3})
\end{equation}
with
\begin{equation}
\alpha^{\,}_{1}:= \sigma^{\,}_{3}\otimes\tau^{\,}_{1},
\qquad
\alpha^{\,}_{2}:= \sigma^{\,}_{3}\otimes\tau^{\,}_{2},
\end{equation}
and
\begin{equation}
\beta^{\,}_{1}:= \sigma^{\,}_{1}\otimes\tau^{\,}_{0},
\qquad
\beta^{\,}_{2}:= \sigma^{\,}_{2}\otimes\tau^{\,}_{0},
\qquad
\beta^{\,}_{3}:= \sigma^{\,}_{3}\otimes\tau^{\,}_{3}.
\end{equation}
\end{subequations}
A second set of Pauli matrices 
$\tau^{\,}_{1}$,
$\tau^{\,}_{2}$,
and
$\tau^{\,}_{1}$
has been introduced, together with the unit $2\times2$
matrices $\sigma^{\,}_{0}$ and $\tau^{\,}_{0}$.
If the triplet of Higgs field $\boldsymbol{\phi}^{\,}_{0}$
is constant throughout two-dimensional space ($\boldsymbol{x}$),
the single-particle spectrum of
Hamiltonian (\ref{eq: def 2D Dirac Hamiltonian})
has a gap controlled by the Higgs components adding in quadrature,
\begin{equation}
\varepsilon^{2}_{0}(\p)
=
\p^{2} + \bm{\phi}^{2}_{0}
=
\p^{2} 
+ 
\phi^{2}_{0\,1}
+ 
\phi^{2}_{0\,2}
+ 
\phi^{2}_{0\,3}.
\end{equation}

The generator of the chiral symmetry of
the Dirac Hamiltonian%
~(\ref{eq: def 2D Dirac Hamiltonian})
is
\begin{subequations}
\begin{equation}
\mathcal{C}^{\,}_{\mathrm{ch}}=
\beta^{\,}_{3}=
\sigma^{\,}_{3}\otimes\tau^{\,}_{3}
\end{equation}
if
\begin{equation}
\phi^{\,}_{3}=0
\end{equation}
\end{subequations}
everywhere in Euclidean space $\boldsymbol{x}\in\mathbb{R}^{2}$.

The operation of time-reversal is implemented by
\begin{equation}
\mathcal{T}^{\,}_{2}\,\mathsf{K}:= 
\sigma^{\,}_{1}\otimes\tau^{\,}_{1}\,
\mathsf{K}
\end{equation}
where $\mathsf{K}$ denotes complex conjugation.
It is a symmetry of the Dirac Hamiltonian%
~(\ref{eq: def 2D Dirac Hamiltonian}) 
for any Higgs configuration 
$\phi^{\,}_{1}$, $\phi^{\,}_{2}$, and $\phi^{\,}_{3}$.

The operation of particle-hole exchange is implemented by
\begin{equation}
\mathcal{C}^{\,}_{\mathrm{ph}}:=
\mathcal{T}^{\,}_{2}\,
\mathsf{K}\,
\mathcal{C}^{\,}_{\mathrm{ch}}=
-
\sigma^{\,}_{2}
\otimes
\tau^{\,}_{2}\,
\mathsf{K}
\end{equation}
and it is only a symmetry of
the Dirac Hamiltonian%
~(\ref{eq: def 2D Dirac Hamiltonian}) 
provided $\phi^{\,}_{3}=0$.

The discovery that this model supports zero modes (Majorana fermions)
was made by Jackiw and Rossi in Ref.~\onlinecite{Jackiw81b}
within an interpretation of Hamiltonian%
~(\ref{eq: def 2D Dirac Hamiltonian})
as a relativistic superconductor.
Weinberg shortly thereafter proved an index theorem
in Ref.~\onlinecite{Weinberg81} 
for these zero modes.
The effect on the induced 
 conserved quasiparticle charge by
a triplet of Higgs fields was investigated by
Jaroszewicz in Ref.~\onlinecite{Jaroszewicz84}
(see also Refs.~\onlinecite{Chen89,Hlousek90,Yakovenko90}).
It was proposed by Hou, Chamon, and Mudry in 
Ref.~\onlinecite{Hou07}
that graphene could realize Hamiltonian%
~(\ref{eq: def 2D Dirac Hamiltonian})
with the Higgs doublet $\phi^{\,}_{1}$ and $\phi^{\,}_{2}$
responsible for a Kekul\'e bond-density-wave instability
and the charge-conjugation-symmetry-breaking $\phi^{\,}_{3}$ responsible for
a charge-density-wave instability
(see also Refs.~\onlinecite{Chamon08a,Chamon08b,Ryu09}).

\subsubsection{
Counting zero modes
              }

We start from the expansion%
~(\ref{eq: master formula before expanding in momenta a})
of the  quasiparticle charge density induced by
a static triplet of Higgs fields $\boldsymbol{\phi}$
which, for a Dirac Hamiltonian,
is exact. 
We compute first the contribution from $n=1$. 
It vanishes.
The adiabatic approximation to 
the quasiparticle charge density 
is in fact given by Eq.~(\ref{eq: n=2 adia appro})
\begin{equation}
\begin{split}
\rho^{\,}_{\mathrm{adia}}(\x)=&\,
\int
\frac{
\mathrm{d}\omega\mathrm{d}^{2}\p
     }
     {
(2\pi)^3
     }
\frac{
8\epsilon^{\,}_{\mathsf{abc}}
\left(\d^{\,}_{1}\phi^{\,}_{\mathsf{a}}\right)(\boldsymbol{x})
\left(\d^{\,}_{2}\phi^{\,}_{\mathsf{b}}\right)(\boldsymbol{x})
\phi^{\,}_{\mathsf{c}}(\boldsymbol{x})
     }
     {
\left[
\omega^{2}
+
\p^{2}
+
\boldsymbol{\phi}^{2}(\boldsymbol{x})
\right]^3
     }
\\
=&\, 
\frac{
\epsilon^{\,}_{\mathsf{abc}}
     }
     {
4\pi
     }
\frac{
\left(\d^{\,}_{1}\phi^{\,}_{\mathsf{a}}\right)(\boldsymbol{x})
\left(\d^{\,}_{2}\phi^{\,}_{\mathsf{b}}\right)(\boldsymbol{x})
\phi^{\,}_{\mathsf{c}}(\boldsymbol{x})
     }
     {
|\boldsymbol{\phi}(\boldsymbol{x})|^{3}
     }.
\end{split}
\end{equation}
For a constant $\phi^{\,}_{3}$, 
the integration over $\x$ gives the adiabatic approximation
to the  conserved quasiparticle charge
\begin{subequations}
\begin{equation}
\begin{split}
Q^{\,}_{\mathrm{adia}}(\phi^{\,}_{3})=&\,
\frac{1}{4\pi}
\int\mathrm{d}\Theta
\int\limits_{0}^{\Delta^{\,}_{0}}\mathrm{d}\rho\,\rho\,
\frac{
\phi^{\,}_{3}
     }
     {
\left(
\rho^{2}
+
\phi^{2}_{3}
\right)^{3/2}
     }
\end{split}
\label{eq: Dirac d=2 number zm}
\end{equation} 
where the parametrization
\begin{equation}
\begin{pmatrix}
\phi_{1}
\\
\phi_{2}
\end{pmatrix}
=
\begin{pmatrix} 
\rho(r)\cos{\Theta(\theta)}
\\
\rho(r)\sin{\Theta(\theta)}
\end{pmatrix}
\end{equation}
\end{subequations}
is assumed for the charge-conjugation-symmetric doublet
of Higgs fields
with $r$ and $\theta$ denoting 
the polar coordinates of $\boldsymbol{x}\in\mathbb{R}^{2}$.
In the limit in which $\phi^{\,}_{3}$ tends to zero, 
the induced charge 
\begin{equation}
Q^{\,}_{\mathrm{adia}}(\phi^{\,}_{3})=
\frac{\mathrm{sgn}(\phi^{\,}_{3})}{2}
\times
\text{winding number in }(\phi^{\,}_{1},\phi^{\,}_{2})
\end{equation} 
follows.
To compute the number of unoccupied zero modes $N$ with
the counting formula%
~(\ref{eq: master counting formula bis}),
we choose the sign of the charge-conjugation-symmetry-breaking
$\phi^{\,}_{3}$ such that it shifts the zero mode in energy
above the chemical potential, i.e., with the opposite sign
to the winding number of the Higgs doublet 
$(\phi^{\,}_{1},\phi^{\,}_{2})$. We then take the limit
$\phi^{\,}_{3}\to 0$. If so, for a unit winding
\begin{eqnarray}
N=1.
\end{eqnarray} 
Observe that the number of zero modes%
~(\ref{eq: Dirac d=2 number zm})
agrees with Weinberg's index theorem
in Ref.~\onlinecite{Weinberg81}
applied to a single vortex with unit vorticity.

\subsubsection{
Chern number
              }
\label{appsubsubsec: Chern number for Dirac if d=2}

We use the notation
$\mathcal{G}:=(\mathrm{i}\omega-\mathcal{H}^{\mathrm{Dirac}}_{d=2})^{-1}$
and compactify both space and the order-parameter space,
\begin{equation}
\boldsymbol{x}\in S^{2},
\qquad
\boldsymbol{\phi}(\boldsymbol{\theta})\in S^{2}\subset\mathbb{R}^{3},
\end{equation}
where $\boldsymbol{\theta}=(\theta^{\,}_{1},\theta^{\,}_{2})$ 
are the spherical coordinates on the two-sphere
$S^{2}\subset\mathbb{R}^{3}$.

With the general manipulations of Appendix%
~\ref{appsubsec: Chern number for Dirac fermions in d-dimensional space},
it is then possible to write
\begin{subequations}
\label{appeq: final Chern d=2 Dirac}
\begin{equation}
\begin{split}
Q^{\,}_{\mathrm{adia}}=&\,
\frac{-\mathrm{i}(2\pi)^{2}}{60}
\int\limits_{K}
\epsilon^{\,}_{\nu^{\,}_{1}\cdots\nu^{\,}_{5}}
\mathrm{tr}^{\,}_{4}
\left(
\mathcal{G}\partial^{\,}_{\nu^{\,}_{1}}\mathcal{G}^{-1}
\cdots
\mathcal{G}\partial^{\,}_{\nu^{\,}_{5}}\mathcal{G}^{-1}
\right)^{\,}_{0}
\end{split}
\label{appeq: final Chern d=2 Dirac a}
\end{equation}
where we have introduced the family of indices $\nu$
\begin{equation}
\nu^{\,}_{1},\cdots,\nu^{\,}_{5}=1,\cdots,5,
\label{appeq: final Chern d=2 Dirac b}
\end{equation}
the momentum
\begin{equation}
K^{\,}_{\nu}=
(\omega,p^{\,}_{1},p^{\,}_{2},\theta^{\,}_{1},\theta^{\,}_{2}),
\label{appeq: final Chern d=2 Dirac c}
\end{equation}
and the domain of integration
\begin{equation}
\int\limits_{K}\equiv
\int\frac{\mathrm{d}\omega}{2\pi}
\int\limits_{\boldsymbol{p}\in S^{2}}
\frac{
\mathrm{d}\Omega^{\,}_{2}(\boldsymbol{p})
     }
     {
(2\pi)^{2}
     }
\int\limits_{\boldsymbol{\theta}\in S^{2}}
\frac{
\mathrm{d}\Omega^{\,}_{2}(\boldsymbol{\theta})
     }
     {
(2\pi)^{2}
     }.
\label{appeq: final Chern d=2 Dirac d}
\end{equation}
\end{subequations}
The ``surface'' element of the sphere $S^{2}$ is here denoted by
$\mathrm{d}\Omega^{\,}_{2}$.
The subscript 0 refers to the semi-classical
Green function~(\ref{eq: def semi classical G}).
Equation~(\ref{appeq: final Chern d=2 Dirac})
is \textit{the second Chern number}.%
~\cite{Qi08} 
It takes integer values only.

\subsection{
Chern number for Dirac fermions in $d$-dimensional space
           }
\label{appsubsec: Chern number for Dirac fermions in d-dimensional space}

To prove Eq.~(\ref{eq: main result on Dirac fermions}),
imagine that we integrate out the Dirac fermions
in the background, not necessarily static, of the Higgs fields
in Eq.%
~(\ref{eq: def Dirac Hamiltonian in any d})
subject to the constraint 
\begin{subequations} 
\label{appeq: relativisctic invariance restricts current}
\begin{equation} 
\boldsymbol{\phi}(\boldsymbol{\theta})\in S^{d}\subset\mathbb{R}^{d+1},
\label{appeq: relativisctic invariance restricts current a}
\end{equation}
where $\boldsymbol{\theta}$ are the polar coordinates of the
$d$-sphere $S^{d}\subset\mathbb{R}^{d+1}$.
The conserved current of the fermionic single-particle
Dirac Hamiltonian must induce a conserved current
$j^{\nu}_{\mathrm{adia}}$ with $\nu=0,1,\cdots,d$
for the Higgs fields. Its time-like component
$j^{0}_{\mathrm{adia}}$
enters in the counting formula%
~(\ref{eq: master counting formula bis}).
Relativistic covariance, current conservation,
dimensional analysis, and the constraint%
~(\ref{appeq: relativisctic invariance restricts current a})
all conspire to bring this current to the form
\begin{equation}
j^{\nu}_{\mathrm{adia}}\propto
\epsilon^{\nu\nu^{\,}_{1}\cdots\nu^{\,}_{d}}
\epsilon^{\,}_{\mathsf{a}^{\,}_{1}\cdots\mathsf{a}^{\,}_{d}\mathsf{a}^{\,}_{d+1}}
\phi^{\,}_{\mathsf{a}^{\,}_{d+1}}
\partial^{\,}_{\nu^{\,}_{d}}
\phi^{\,}_{\mathsf{a}^{\,}_{d}}
\cdots
\partial^{\,}_{\nu^{\,}_{1}}
\phi^{\,}_{\mathsf{a}^{\,}_{1}}.
\label{appeq: relativisctic invariance restricts current b}
\end{equation}
Here, summation convention over the repeated indices
\begin{equation}
\begin{split}
&
\nu,\nu^{\,}_{1},\cdots,\nu^{\,}_{d}=
0,1,\cdots,d,
\\
&
\mathsf{a}^{\,}_{1},\cdots,\mathsf{a}^{\,}_{d},\mathsf{a}^{\,}_{d+1}=
d+1,\cdots,2d+1,
\end{split}
\label{appeq: relativisctic invariance restricts current c}
\end{equation}
\end{subequations}
is implied. (Compared to our convention in
the definition%
~(\ref{eq: def Dirac Hamiltonian in any d})
of the Dirac Hamiltonian,
we have shifted the values taken by
the family of indices $\mathsf{a}=d+1,\cdots,2d+1$
to stress that it differs from the family of indices 
$i=1,\cdots,d$.)

If we compare Eq.%
~(\ref{appeq: relativisctic invariance restricts current})
with the gradient expansion%
~(\ref{eq: leading n=d order if no double derivative}),
we deduce that the leading non-vanishing
contribution to the gradient expansion%
~(\ref{eq: leading n=d order if no double derivative})
must be of order $n=d$ and, for a static
Higgs background, given by
\begin{subequations}
\label{appeq: gradient expansion compared to induced current} 
\begin{equation}
j^{0}_{\mathrm{adia}}(\boldsymbol{x})=
(-\mathrm{i})^{d}
\mathcal{I}^{\;}_{
i^{\,}_{d}\mathsf{a}^{\,}_{d}\cdots i^{\,}_{1}\mathsf{a}^{\,}_{1}
                }
\left(\partial^{\,}_{i^{\,}_{d}}\phi^{\,}_{\mathsf{a}^{\,}_{d}}\right)
\cdots
\left(\partial^{\,}_{i^{\,}_{1}}\phi^{\,}_{\mathsf{a}^{\,}_{1}}\right)
(\boldsymbol{x}).
\label{appeq: gradient expansion compared to induced current a} 
\end{equation}
The summation convention over the two distinct families of indices
\begin{equation}
\begin{split}
&
i^{\,}_{1},\cdots,i^{\,}_{d}=1,\cdots,d,
\\
&
\mathsf{a}^{\,}_{1},\cdots,\mathsf{a}^{\,}_{d}=d+1,\cdots,2d+1,
\end{split}
\label{appeq: gradient expansion compared to induced current b} 
\end{equation}
is implied and the expansion coefficients are
\begin{equation}
\begin{split}
\mathcal{I}^{\;}_{
i^{\,}_{d}\mathsf{a}^{\,}_{d}\cdots i^{\,}_{1}\mathsf{a}^{\,}_{1}
                }:=&
-\mathrm{i}
\int\frac{\mathrm{d}\omega}{2\pi}
\int\limits_{\boldsymbol{p}}
\mathrm{tr}^{\,}_{R}
\Bigg[
\left(
\mathcal{G}\frac{\partial\mathcal{G}^{-1}}{\partial p^{\,}_{i^{\,}_{d}}}
\mathcal{G}\frac{\partial\mathcal{G}^{-1}}{\partial \phi^{\,}_{\mathsf{a}^{\,}_{d}}}
\right)
\cdots
\\
&
\cdots
\left(
\mathcal{G}\frac{\partial\mathcal{G}^{-1}}{\partial p^{\,}_{i^{\,}_{1}}}
\mathcal{G}\frac{\partial\mathcal{G}^{-1}}{\partial \phi^{\,}_{\mathsf{a}^{\,}_{1}}}
\right)
\left(
\mathcal{G}\frac{\partial\mathcal{G}^{-1}}{\partial\omega}
\right)
\Bigg]^{\,}_{0}(\omega,\boldsymbol{p}).
\end{split}
\label{appeq: gradient expansion compared to induced current c} 
\end{equation}
\end{subequations}
Any permutation of the indices on the left-hand side is defined by
the same permutation of the differentials in the trace of the right-hand
side.

We are first going to prove that
\begin{subequations}
\label{appeq: intermediary step in proof generic Dirac}
\begin{equation}
\begin{split}
Q^{\,}_{\mathrm{adia}}:=&
\int\limits_{S^{d}}\mathrm{d}^{d}\boldsymbol{x}\,
j^{0}_{\mathrm{adia}}(\boldsymbol{x})
\\
=&
\frac{
(-)^{d(d-1)/2}(-\mathrm{i})^{d}
     }
     {
d!
     }
\!\!\int\limits_{\boldsymbol{\theta}\in S^{d}}\!\!
\epsilon^{\;}_{
i^{\,}_{d}\mathsf{b}^{\,}_{d}\cdots i^{\,}_{1}\mathsf{b}^{\,}_{1}
              }
\mathcal{J}^{\;}_{
i^{\,}_{d}\mathsf{b}^{\,}_{d}\cdots i^{\,}_{1}\mathsf{b}^{\,}_{1}
                 }
(\boldsymbol{\theta})
\end{split}
\label{appeq: intermediary step in proof generic Dirac a}
\end{equation}
where the target space $S^{d}\subset\mathbb{R}^{d+1}$ 
of the order-parameter is parametrized by the
$d$-independent spherical coordinates
\begin{equation}
\boldsymbol{\theta}\equiv
(\theta^{\,}_{d+1},\cdots,\theta^{\,}_{2d}),
\label{appeq: intermediary step in proof generic Dirac b}
\end{equation}
the integral
\begin{equation}
\int\limits_{\boldsymbol{\theta}\in S^{d}}\equiv
\int\limits_{S^{d}}
\mathrm{d}\Omega^{\,}_{d}
\end{equation}
with $\mathrm{d}\Omega^{\,}_{d}$
the ``surface'' element of the $d$-sphere,
and 
\begin{equation}
\begin{split}
\mathcal{J}^{\,}_{
i^{\,}_{d}\mathsf{b}^{\,}_{d}\cdots i^{\,}_{1}\mathsf{b}^{\,}_{1}
                }:=&
-\mathrm{i}
\int\frac{\mathrm{d}\omega}{2\pi}
\int\limits_{\boldsymbol{p}}
\mathrm{tr}^{\,}_{R}
\Bigg[
\left(
\mathcal{G}\frac{\partial\mathcal{G}^{-1}}{\partial p^{\,}_{i^{\,}_{d}}}
\mathcal{G}\frac{\partial\mathcal{G}^{-1}}{\partial \theta^{\,}_{\mathsf{b}^{\,}_{d}}}
\right)
\cdots
\\
&
\cdots
\left(
\mathcal{G}\frac{\partial\mathcal{G}^{-1}}{\partial p^{\,}_{i^{\,}_{1}}}
\mathcal{G}\frac{\partial\mathcal{G}^{-1}}{\partial \theta^{\,}_{\mathsf{b}^{\,}_{1}}}
\right)
\left(
\mathcal{G}\frac{\partial\mathcal{G}^{-1}}{\partial\omega}
\right)
\Bigg]^{\,}_{0}(\omega,\boldsymbol{p}).
\end{split}
\label{appeq: intermediary step in proof generic Dirac c}
\end{equation}
\end{subequations}
Again any permutation of the indices on the left-hand side is defined by
the same permutation of the differentials in the trace of the right-hand
side.

\begin{proof}
To evaluate the trace in the integrand%
~(\ref{appeq: gradient expansion compared to induced current c}),
the semi-classical
Green functions are re-massaged so as to bring
all the Dirac matrices in the numerator,
\begin{equation}
\begin{split}
\mathcal{G}^{\,}_{\text{s-c}}
(\omega,\boldsymbol{p},\boldsymbol{x}):=&\,
\frac{
1
     }
     {
\mathrm{i}\omega
-
\Gamma^{\,}_{i}p^{\,}_{i}
-
\Gamma^{\,}_{\mathsf{a}}\phi^{\,}_{\mathsf{a}}(\boldsymbol{x})
     }
\\
=&\,
-
\frac{
\mathrm{i}\omega
+
\Gamma^{\,}_{i}p^{\,}_{i}
+
\Gamma^{\,}_{\mathsf{a}}\phi^{\,}_{\mathsf{a}}(\boldsymbol{x})
     }
     {
\omega^{2}
+
\boldsymbol{p}^{2}
+
\boldsymbol{\phi}^{2}(\boldsymbol{x})
     }.
\end{split}
\label{appeq: semi classical Dirac green}
\end{equation}
Multiplying out all Green functions in the trace
from the integrand in Eq.%
~(\ref{appeq: gradient expansion compared to induced current c})
yields in the numerator
terms made of the product from
$2d$ $\Gamma$-matrices,
$2d+1$ $\Gamma$-matrices, ...,
$2d+2j$ $\Gamma$-matrices,
$2d+2j+1$ $\Gamma$-matrices, ...,
$2d+2d$ $\Gamma$-matrices,
and $2d+2d+1$ $\Gamma$-matrices.
Any trace over an even number $2d+2j$ of
$\Gamma$-matrices is odd under
$\omega\to-\omega$ since it comes multiplied by the power
$\omega^{2d+1-2j}$ in the numerator. 
Such a trace does not contribute to the
$\omega$ integration
since the denominator is an even function of $\omega$. 
Any trace over an odd number $2d+2j+1$ of
$\Gamma$-matrices is even under
$\omega\to-\omega$ since it comes multiplied by the power
$\omega^{2d+1-2j-1}$. Such a trace can only be nonvanishing if
$\Gamma^{\,}_{1}$,$\cdots,$$\Gamma^{\,}_{2d+1}$
all appear in the trace and all an odd number of times.%
~\cite{Zinnjustin-textbook}
For such traces,
the Clifford algebra delivers another key identity in that
\begin{equation}
\mathrm{tr}^{\,}_{R} 
\left(
\Gamma^{\,}_{2d+1}
\Gamma^{\,}_{\mu^{\,}_{1}}
\cdots
\Gamma^{\,}_{\mu^{\,}_{2d}}
\right)
=
R\,(+\mathrm{i})^d
\epsilon^{\,}_{\mu^{\,}_{1}\cdots\mu^{\,}_{2d}}
\label{appeq: master trace formula for Dirac}
\end{equation}
if $\mu^{\,}_{j}=1,\cdots,2d$ for $j=1,\cdots,2d$.
Here, $R=2^{d}$ and $\Gamma^{\,}_{2d+1}$
was defined in Eq.%
~(\ref{eq: def Gamma 2d+1}).
The coefficients%
~(\ref{appeq: gradient expansion compared to induced current c})
inherit the antisymmetry of Eq.%
~(\ref{appeq: master trace formula for Dirac})
in that they are fully antisymmetric under any exchange
of the indices 
(\ref{appeq: gradient expansion compared to induced current b}).

We need to overcome the fact that the ranges of
$i$ and $\mathsf{a}$ are unequal in cardinality. 
To this end, 
we change variables on the target space and introduce the
spherical coordinates%
~(\ref{appeq: intermediary step in proof generic Dirac b})
of the target space $S^{d}\subset\mathbb{R}^{d+1}$. 
The adiabatic approximation%
~(\ref{appeq: gradient expansion compared to induced current a})
to the quasiparticle charge density becomes
\begin{equation}
j^{0}_{\mathrm{adia}}(\boldsymbol{x})=
(-\mathrm{i})^{d}
\mathcal{J}^{\;}_{
i^{\,}_{d}\mathsf{b}^{\,}_{d}\cdots i^{\,}_{1}\mathsf{b}^{\,}_{1}
                }
\left(\partial^{\,}_{i^{\,}_{d}}\theta^{\,}_{\mathsf{b}^{\,}_{d}}\right)
\cdots
\left(\partial^{\,}_{i^{\,}_{1}}\theta^{\,}_{\mathsf{b}^{\,}_{1}}\right)
(\boldsymbol{x}).
\label{appeq: gradient expansion compared to induced current if theta}
\end{equation}
Now, it is the summation convention over the two distinct families of indices
\begin{equation}
\begin{split}
&
i^{\,}_{1},\cdots,i^{\,}_{d}=1,\cdots,d,
\\
&
\mathsf{b}^{\,}_{1},\cdots,\mathsf{b}^{\,}_{d}=d+1,\cdots,2d,
\end{split}
\label{appeq: range indices if theta} 
\end{equation}
that replaces%
~(\ref{appeq: gradient expansion compared to induced current b}),
whereby the expansion coefficients%
~(\ref{appeq: gradient expansion compared to induced current if theta})
are related to the expansion coefficients%
~(\ref{appeq: gradient expansion compared to induced current c})
through the chain rule for differentiation, i.e.,
\begin{equation}
\mathcal{J}^{\;}_{
i^{\,}_{d}\mathsf{b}^{\,}_{d}\cdots i^{\,}_{1}\mathsf{b}^{\,}_{1}
                }=
\mathcal{I}^{\;}_{
i^{\,}_{d}\mathsf{a}^{\,}_{d}\cdots i^{\,}_{1}\mathsf{a}^{\,}_{1}
                }
\left(
\frac{
\partial
\phi^{\,}_{\mathsf{a}^{\,}_{d}}
     }
     {
\partial
\theta^{\,}_{\mathsf{b}^{\,}_{d}}
     }
\right)
\cdots
\left(
\frac{
\partial
\phi^{\,}_{\mathsf{a}^{\,}_{1}}
     }
     {
\partial
\theta^{\,}_{\mathsf{b}^{\,}_{1}}
     }
\right).
\end{equation}
Here, any permutation of the indices of $\mathcal{J}$
on the left-hand side is defined by
the same permutation on the indices of $\mathcal{I}$ 
on the right-hand side.

By linearity, the antisymmetry%
~(\ref{appeq: master trace formula for Dirac})
thus carries over to 
\begin{equation}
\mathcal{J}^{\;}_{
i^{\,}_{d}\mathsf{b}^{\,}_{d}\cdots i^{\,}_{1}\mathsf{b}^{\,}_{1}
                }=
\mathcal{N}^{\,}_{d}\,
\epsilon^{\;}_{i^{\,}_{d}\mathsf{b}^{\,}_{d}\cdots i^{\,}_{1}\mathsf{b}^{\,}_{1}}
\end{equation} 
where $\mathcal{N}^{\,}_{d}$ is a normalization constant.
We will not need the explicit dependence of the 
normalization $\mathcal{N}^{\,}_{d}$.
We will only make use of the fact that it obeys the identity
\begin{equation}
\mathcal{N}^{\,}_{d}(\boldsymbol{\theta})=
\frac{1}{(d!)^{2}}
\epsilon^{\;}_{
i^{\,}_{d}\mathsf{b}^{\,}_{d}\cdots i^{\,}_{1}\mathsf{b}^{\,}_{1}
              }
\mathcal{J}^{\;}_{
i^{\,}_{d}\mathsf{b}^{\,}_{d}\cdots i^{\,}_{1}\mathsf{b}^{\,}_{1}
                 }
(\boldsymbol{\theta}).
\label{appeq: auxiliary identity for Normalization}
\end{equation}
This follows from contracting the Levi-Civita antisymmetric tensor
$\epsilon^{\;}_{i^{\,}_{d}\mathsf{b}^{\,}_{d}\cdots i^{\,}_{1}\mathsf{b}^{\,}_{1}}$
with itself and observing that the two sets of indices $i$ and $\mathsf{b}$
run over $d$ distinct values each.

The Jacobian of the map from $\boldsymbol{x}\in S^{d}$ to
$\boldsymbol{\theta}\in S^{d}\subset\mathbb{R}^{d+1}$ is
\begin{equation}
\begin{split}
\left|
\frac{\partial\boldsymbol{\theta}}{\partial\boldsymbol{x}}
\right|=&\,
\epsilon^{\,}_{i^{\,}_{d}\cdots i^{\,}_{1}}
\partial^{\,}_{i^{\,}_{d}}\theta^{\,}_{2d}
\cdots
\partial^{\,}_{i^{\,}_{1}}\theta^{\,}_{d+1}
\\
=&\,
\frac{1}{d!}
\epsilon^{\,}_{i^{\,}_{d}\cdots i^{\,}_{1}}
\epsilon^{\,}_{\mathsf{b}^{\,}_{d}\cdots\mathsf{b}^{\,}_{1}}
\partial^{\,}_{i^{\,}_{d}}\theta^{\,}_{\mathsf{b}^{\,}_{d}}
\cdots
\partial^{\,}_{i^{\,}_{1}}\theta^{\,}_{\mathsf{b}^{\,}_{1}}
\\
=&\,
\frac{1}{d!}
\epsilon^{\,}_{i^{\,}_{d}\cdots i^{\,}_{1}\mathsf{b}^{\,}_{d}\cdots\mathsf{b}^{\,}_{1}}
\partial^{\,}_{i^{\,}_{d}}\theta^{\,}_{\mathsf{b}^{\,}_{d}}
\cdots
\partial^{\,}_{i^{\,}_{1}}\theta^{\,}_{\mathsf{b}^{\,}_{1}}
\\
=&\,
\frac{(-)^{d(d-1)/2}}{d!}
\epsilon^{\,}_{i^{\,}_{d}\mathsf{b}^{\,}_{d}\cdots i^{\,}_{1}\mathsf{b}^{\,}_{1}}
\partial^{\,}_{i^{\,}_{d}}\theta^{\,}_{\mathsf{b}^{\,}_{d}}
\cdots
\partial^{\,}_{i^{\,}_{1}}\theta^{\,}_{\mathsf{b}^{\,}_{1}}.
\end{split}
\end{equation}
The conserved quasiparticle charge then becomes
\begin{equation}
\begin{split}
Q^{\,}_{\mathrm{adia}}:=&
\int\limits_{\boldsymbol{x}\in S^{d}}\,
j^{0}_{\mathrm{adia}}
\\
=&
(-\mathrm{i})^{d+1}
\int\limits_{\boldsymbol{x}\in S^{d}}\,
\mathcal{J}^{\;}_{
i^{\,}_{d}\mathsf{b}^{\,}_{d}\cdots i^{\,}_{1}\mathsf{b}^{\,}_{1}
                }
\partial^{\,}_{i^{\,}_{d}}\theta^{\,}_{\mathsf{b}^{\,}_{d}}
\cdots
\partial^{\,}_{i^{\,}_{1}}\theta^{\,}_{\mathsf{b}^{\,}_{1}}
\\
=&
(-\mathrm{i})^{d+1}
\int\limits_{\boldsymbol{x}\in S^{d}}\,
\mathcal{N}^{\,}_{d}\,
\epsilon^{\;}_{i^{\,}_{d}\mathsf{b}^{\,}_{d}\cdots i^{\,}_{1}\mathsf{b}^{\,}_{1}}
\partial^{\,}_{i^{\,}_{d}}\theta^{\,}_{\mathsf{b}^{\,}_{d}}
\cdots
\partial^{\,}_{i^{\,}_{1}}\theta^{\,}_{\mathsf{b}^{\,}_{1}}
\\
=&
(-\mathrm{i})^{d+1}
(-)^{d(d-1)/2}d!
\int\limits_{\boldsymbol{\theta}\in S^{d}}
\mathcal{N}^{\,}_{d}(\boldsymbol{\theta})
\\
=&\,
\frac{
(-)^{d(d-1)/2}(-\mathrm{i})^{d+1}
     }
     {
d!
     }
\!\!\int\limits_{\boldsymbol{\theta}\in S^{d}}\!\!
\epsilon^{\;}_{
i^{\,}_{d}\mathsf{b}^{\,}_{d}\cdots i^{\,}_{1}\mathsf{b}^{\,}_{1}
              }
\mathcal{J}^{\;}_{
i^{\,}_{d}\mathsf{b}^{\,}_{d}\cdots i^{\,}_{1}\mathsf{b}^{\,}_{1}
                 }
(\boldsymbol{\theta})
\end{split}
\end{equation}
with the help of Eq.%
~(\ref{appeq: auxiliary identity for Normalization}) 
to reach the last equality.

\end{proof}

For convenience, we introduce another family of indices $\mu$
through
\begin{equation}
\begin{split}
&
\mu^{\,}_{1},\cdots,\mu^{\,}_{2d}=1,\cdots,2d,
\\
&
i^{\,}_{1},\cdots,i^{\,}_{d}=1,\cdots,d,
\\
&
\mathsf{b}^{\,}_{1},\cdots,\mathsf{b}^{\,}_{d}=d+1,\cdots,2d.
\end{split}
\label{appeq: gradient expansion compared to induced current b bis} 
\end{equation}
For any tensor $\mathcal{J}^{\,}_{\mu^{\,}_{1}\cdots\mu^{\,}_{2d}}$
that reduces to the fully antisymmetric tensor
$
\mathcal{J}^{\;}_{
i^{\,}_{d}\mathsf{b}^{\,}_{d}\cdots i^{\,}_{1}\mathsf{b}^{\,}_{1}
                }
$,
we have the identity
\begin{equation}
\epsilon^{\,}_{\mu^{\,}_{1}\cdots\mu^{\,}_{2d}}
\mathcal{J}^{\,}_{\mu^{\,}_{1}\cdots\mu^{\,}_{2d}}=
\frac{
(2d)!
     }
     {
(d!)^{2}
     }\,
\epsilon^{\;}_{i^{\,}_{d}\mathsf{b}^{\,}_{d}\cdots 
               i^{\,}_{1}\mathsf{b}^{\,}_{1}}
\mathcal{J}^{\;}_{
i^{\,}_{d}\mathsf{b}^{\,}_{d}\cdots i^{\,}_{1}\mathsf{b}^{\,}_{1}
                 },
\end{equation}
since the contraction on the left-hand side of this equation yields
a combinatorial factor of $(2d)!$ whereas the contraction on the
right-hand side yields a combinatorial factor of $d!\times d!$,
i.e., one $d!$ for the family $i$ of indices and another $d!$
for the distinct family $\mathsf{b}$ of indices.
In particular, we can choose
\begin{equation}
\mathcal{J}^{\,}_{\mu^{\,}_{1}\cdots\mu^{\,}_{2d}}:=\!
\int\limits_{\omega}\!
\int\limits_{\boldsymbol{p}\in S^{d}}\!\!\!
\mathrm{tr}^{\,}_{R}
\left(
\mathcal{G}\partial^{\,}_{\mu^{\,}_{1}}\mathcal{G}^{-1}
\cdots
\mathcal{G}\partial^{\,}_{\mu^{\,}_{2d}}\mathcal{G}^{-1}
\mathcal{G}\partial^{\,}_{\omega}\mathcal{G}^{-1}
\right)^{\,}_{0}
\end{equation}
where the subscript 0 refers to the semi-classical
Green function~(\ref{appeq: semi classical Dirac green}).

At last, we add the non-compact imaginary-time label
with the introduction of the family $\nu$ of indices,
\begin{equation}
\begin{split}
&
\nu^{\,}_{1},\cdots,\nu^{\,}_{2d+1}=0,1,\cdots,2d,
\\
&
\mu^{\,}_{1},\cdots,\mu^{\,}_{2d}=1,\cdots,2d,
\\
&
i^{\,}_{1},\cdots,i^{\,}_{d}=1,\cdots,d,
\\
&
\mathsf{b}^{\,}_{1},\cdots,\mathsf{b}^{\,}_{d}=d+1,\cdots,2d.
\end{split}
\label{appeq: gradient expansion compared to induced current b bis bis} 
\end{equation}
Define
\begin{subequations}
\label{appeq: proof d chen number} 
\begin{equation}
\begin{split}
&
K^{\,}_{\nu}:=
(\omega,p^{\,}_{1},\cdots,p^{\,}_{d},\theta^{\,}_{d+1},\cdots,\theta^{\,}_{2d}),
\\
&
\mathcal{J}^{\,}_{\nu^{\,}_{1}\cdots\nu^{\,}_{2d+1}}:=
\int\limits_{\omega}
\int\limits_{\boldsymbol{p}\in S^{d}}
\mathrm{tr}^{\,}_{R}
\left(
\mathcal{G}\partial^{\,}_{\nu^{\,}_{1}}\mathcal{G}^{-1}
\cdots
\mathcal{G}\partial^{\,}_{\nu^{\,}_{2d+1}}\mathcal{G}^{-1}
\right)^{\,}_{0},
\\
&
\int\limits_{K}\equiv
\int\limits_{\omega}
\int\limits_{\boldsymbol{p}\in S^{d}}
\int\limits_{\boldsymbol{\theta}\in S^{d}}\equiv
\int\frac{\mathrm{d}\omega}{2\pi}
\int\limits_{S^{d}}
\frac{
\mathrm{d}\Omega^{\,}_{d}(\boldsymbol{p})
     }
     {
(2\pi)^{d}
     }
\int\limits_{S^{d}}
\frac{
\mathrm{d}\Omega^{\,}_{d}(\boldsymbol{\theta})
     }
     {
(2\pi)^{d}
     }.
\end{split}
\label{appeq: def integral in spherical coordinateS}
\end{equation}
The ``surface'' element of $S^{d}$  is here denoted by
$\mathrm{d}\Omega^{\,}_{d}$.
It follows that 
\begin{equation}
\begin{split}
Q^{\,}_{\mathrm{adia}}=&\,
\frac{
(-)^{d(d-1)/2}(-\mathrm{i})^{d+1}d!
     }
     {
(2d+1)!
     }
(2\pi)^{d}
\\
&\times
\int\limits_{K}
\epsilon^{\;}_{\nu^{\,}_{1}\cdots\nu^{\,}_{2d+1}}
\mathrm{tr}^{\,}_{R}
\left(
\mathcal{G}\partial^{\,}_{\nu^{\,}_{1}}\mathcal{G}^{-1}
\cdots
\mathcal{G}\partial^{\,}_{\nu^{\,}_{2d+1}}\mathcal{G}^{-1}
\right)^{\,}_{0}
\end{split}
\end{equation}
\end{subequations}
thereby completing the proof of Eq.%
~(\ref{eq: main result on Dirac fermions}).
This is \textit{the $d$-th Chern number} 
in $(2d+1)$-dimensional Euclidean space and time.

\vskip 4 true cm


\begin{thebibliography}{99}
\bibitem{footnote: comment on spin-wave app}
This additive decomposition is very similar to the spin-wave
approximation. It is only valid locally and fails to capture
the non-trivial topology of a defect such as a vortex.
However, we will remedy this difficulty by the introduction
of an additional charge-symmetry-breaking component to
the Higgs multiplet for which this additive decomposition
becomes meaningful globally.
\bibitem{footnote: semi-classical gradient expansion}
The vanishing term 
$
0=
\mathcal{G}^{\,}_{0}(\omega)
-
\mathcal{G}^{\,}_{0}(\omega)
$
that arises from the order $n=0$ in the geometric series%
~(\ref{eq: geometric expansion Green function diff}) 
need not vanish anymore after performing the semi-classical 
substitution by which 
$
\mathcal{G}^{\,}_{0}(\omega)
-
\mathcal{G}^{\,}_{0}(\omega)
\to
\mathcal{G}^{\,}_{\text{s-c}}(\omega)
-
\mathcal{G}^{\,}_{0}(\omega)
$.
However, the term 
$\int\limits_{\omega,\boldsymbol{p}}
\mathrm{tr}^{\,}_{R}
\left[
\mathcal{G}^{\,}_{\text{s-c}}(\omega,\boldsymbol{p},\boldsymbol{x})
-
\mathcal{G}^{\,}_{0}(\omega,\boldsymbol{p})
\right]
$
does not contribute to the 
quasiparticle charge density provided
that the traces $\mathrm{tr}^{\,}_{R}$
over the Euclidean Green functions
(\ref{eq: def mathcal{G}{0}})
and 
(\ref{eq: def semi classical G})
both yield odd functions of $\omega$. 
This is verified as a consequence of
$
\mathcal{H}^{2}
(\boldsymbol{p},\boldsymbol{\phi}^{\,}_{0})$
being proportional to the unit matrix and
$\mathrm{tr}^{\,}_{R}\,
\mathcal{H}
(\boldsymbol{p},\boldsymbol{\phi}^{\,}_{0})=0$
for all the Hamiltonians studied in this paper.
\bibitem{footnote: gradient expansion for Dirac}
The Taylor expansion of each Green function in Eq.%
~(\ref{eq: master formula before expanding in momenta c})
is performed to the $d$-th order. However, the identity
\begin{equation}
\begin{split}
\left(
\partial^{n}_{i^{\,}_{1}\cdots i^{\,}_{n}}\mathcal{G}
\right)^{\,}_{0}=&\,
(-)^{n} 
\left(
\mathcal{G}
\partial^{\,}_{i^{\,}_{1}}\mathcal{G}^{-1}
\cdots
\mathcal{G}
\partial^{\,}_{i^{\,}_{n}}\mathcal{G}^{-1}
\mathcal{G}
\right)^{\,}_{0}
\\
& 
+ 
\hbox{
all permutations of $i^{\,}_{1},\cdots,i^{\,}_{n}$
     }
\end{split}
\nonumber
\end{equation}
(the Leibnitz rule for the differentiation of 
$\mathcal{G}\mathcal{G}^{-1}$ when
$\mathcal{G}^{-1}$ is linear in momentum)
in combination with the fact that the trace%
~(\ref{appeq: master trace formula for Dirac})
over the Dirac matrices is fully antisymmetric 
can be used to show that
only the linear expansion of each Green function
in Eq.%
~(\ref{eq: master formula before expanding in momenta c})
matters.
\bibitem{footnote: def m sphere}
The $m$-sphere $S^{m}$ is the set of point in $(m+1)$-dimensional 
Euclidean space a unit distance away from the origin.
\bibitem{footnote-index-thm}
Weinberg shows in Ref.~\onlinecite{Weinberg81}
that the index of the elliptic operators
$Q$ and $Q^{\dag}$ making up the two chiral blocks of 
the Jackiw-Rossi Dirac Hamiltonian is a surface integral
without relying on any approximation. In this
paper, we perform first a gradient expansion and then
show that the number of unoccupied zero modes is
a surface integral.
\end{thebibliography}
\end{document}